\documentclass[iop]{emulateapj}

\usepackage{xcolor}

\newcommand{\kms}{\rm ~ km\,s^{-1}}
\newcommand{\chandra}{{\em Chandra~}}
\newcommand{\xmm}{{\em XMM-Newton~}}
\newcommand{\swift}{{\em Swift-XRT~}}
\newcommand{\nustar}{{\em NuSTAR~}}

\newcommand{\sn}{SN 2010jl~}
\newcommand{\snb}{SN 2010jl}
\newcommand{\ugc}{UGC 5189A~}

\newcommand{\Msun}{~M_\odot}

\slugcomment{To appear in ApJ}

\shorttitle{Circumstellar interaction in SN 2010jl}
\shortauthors{Chandra et al.}

\begin{document}


\title{X-RAY AND RADIO EMISSION FROM TYPE IIn SUPERNOVA SN 2010jl}

\author{Poonam\,Chandra\altaffilmark{1},
Roger\,A.\,Chevalier\altaffilmark{2},
Nikolai\,Chugai\altaffilmark{3},
Claes\,Fransson\altaffilmark{4}, and
Alicia\,M.\,Soderberg\altaffilmark{5}}

\altaffiltext{1}{
National Centre for Radio Astrophysics, Tata Institute of Fundamental Research, Pune University Campus, Pune 411 007, INDIA, 
\email{poonam@ncra.tifr.res.in}}
\altaffiltext{2}{Department of Astronomy, University of Virginia, P.O. Box 400325,
Charlottesville, VA 22904-4325}
\altaffiltext{3}{Institute of Astronomy of Russian Academy of Sciences, Pyatnitskaya St. 48,
109017 Moscow, Russia}
\altaffiltext{4}{Oskar Klein Centre, Department of Astronomy, Stockholm University, AlbaNova, SE-106 91 Stockholm,
Sweden}
\altaffiltext{5}{Smithsonian Astrophysical Observatory, 60 Garden St., MS-20, Cambridge,
MA 02138}

\begin{abstract}

We present all X-ray and radio observations of the Type IIn supernova
SN 2010jl.  The X-ray observations cover a period  
up to day 1500 with \chandra, \xmm, \nustar, and \swift.
The  \chandra observations after 2012 June, the \xmm observation
in 2013 November, and most of the \swift observations until 2014
December are presented for the first time.
All the spectra can be fitted by an absorbed hot thermal model except for \chandra spectra on 2011 October and 2012 June when an additional component is needed. Although the origin of this component is uncertain,  it is spatially coincident with the supernova and occurs when there are 
 changes to the supernova spectrum in the energy range close to that of the extra component, indicating that  the emission is related to the supernova. The X-ray light curve shows an initial plateau followed by a steep drop starting at day $\sim 300$. We attribute the drop to a decrease in the circumstellar density. The column density to the X-ray emission drops rapidly with time, showing that the absorption is in the vicinity of the supernova. We also present Very Large Array radio observations of SN 2010jl. Radio emission was detected from \sn from day 570 onwards. The radio light curves and spectra suggest that
the radio luminosity was close to its maximum at the first detection.  
The velocity of the shocked ejecta derived assuming synchrotron self absorption is much less than that estimated from the optical and X-ray observations,  suggesting that free-free absorption dominates.

\end{abstract}

\keywords{
circumstellar matter --- 
stars: mass-loss ---
radiation mechanisms: non-thermal --- 
radio continuum: general --- 
Supernovae: Individual (SN 2010jl)--- 
X-rays: general}
 
\section{INTRODUCTION}
\label{sec:intro}

Type IIn (narrow line) supernovae (SNe) are 
characterized by narrow emission lines atop broad wings, 
slow evolution, and
a blue continuum at early times \citep{schlegel90}.
Their high   H$\alpha$
and bolometric luminosities can be explained by the 
shock interaction of supernova (SN) ejecta 
with a dense circumstellar medium \citep[CSM; ][]{chugai1990}.
The shock waves accompanying the circumstellar interaction 
heat gas to X-ray emitting
temperatures and accelerate particles to relativistic energies, 
giving rise to
radio synchrotron emission.
Indeed, Type IIn supernovae (SNe IIn) are among the most luminous radio and X-ray SNe,
e.g., SN 1986J \citep{bp92}, SN 1988Z \citep{ft96},
 SN 1995N \citep{chandra05}, and SN 2006jd \citep{chandra12b}.

Although high CSM densities should enable radio and X-ray                
emission,  few SNe IIn are detected in these bands.
Amongst the detected ones, the X-ray and radio light curves of these SNe
 cover a range of luminosities
\citep[e.g.,][]{dg12}.
\citet{vandyk96} carried out a study of 10 SNe IIn with the Very Large Array (VLA), but did not detect  radio emission from any of them.   
Type IIn SN 1998S was not 
particularly luminous at radio and
X-ray wavelengths \citep{pooley02},
which can be attributed to a relatively
low CSM density.
However, SN 2006gy was very luminous at optical 
wavelengths, implying a very high
CSM density, but was not luminous at 
X-ray wavelengths \citep{ofek07,smith12}.
The lack of X-ray emission here can be attributed 
to mechanisms that suppress the
X-ray emission at high density, including 
photoelectric absorption, inverse Compton losses
of hot shocked electrons, and Compton cooling in 
the slow wind \citep{chevalier12, svirski12}.
The X-ray luminosity of a SN may initially increase with CSM density, but eventually turns over                  
because of a variety of effects that suppress X-ray emission.

In this paper, we discuss X-ray and radio observations of SN 2010jl, which may be close to the case of a
maximum X-ray luminosity.
 SN 2010jl was discovered 
with a   magnitude of 13.5  in unfiltered CCD images
with a 0.40-m
reflector at Portal, AZ, U.S.A.
 on 2010 November 3  \citep{np10} and
brightened to mag 12.9 over the next day,
showing that it was discovered at an
early phase.
 SN 2010jl is at a position $\alpha=09^h42^m53^s.337$, $\delta=+09^o29'42.''13$ 
(J2000) \citep{ofek14}, associated with a galaxy UGC
5189A at a distance of 49 Mpc ($z=0.0107$),
implying that SN 2010jl belongs to the class of luminous SNe IIn with an
absolute visual magnitude  $M_v<-20$.
Pre-discovery observations indicate an explosion date in  early 2010  October 
\citep{stoll11}. \citet{ofek14} argue for an explosion date around $15-25$
days before I-band maximum, i.e. around 
JD 2,455,469--2,455,479, or 2010 September~29--October~9.
We assume 2010 October 1
to be the explosion date for \sn throughout this paper.
 \citet{stoll11} found that the host galaxy
for \sn is of low metallicity, supporting
the emerging trend that luminous supernovae (SNe)
occur in low metallicity environments. 
They determined the metallicity $Z$ of the
SN region to be $\lesssim0.3~  Z_\odot$.
We take this upper limit as the metallicity of the gas in the galaxy.

After the X-ray Telescope (XRT) on-board {\em Swift}  detected X-rays
from SN 2010jl on 2010 November $5.0-5.8$ \citep{immler10}, we
triggered our {\em Chandra} Target of Opportunity (ToO) 
observing program in 2010 December and 2011 October. 
These observations were
presented in \citet{chandra12a},  in which a very rapid evolution of the
column density was reported. \citet{chandra12a} also reported a constant X-ray flux of the
SN, consistent  with optical wavelengths
where it  displayed a  flat light curve early on \citep{zhang12,ofek14,fec+14}.
While the peak R-band luminosity of SN 2010jl is smaller than the 
super-luminous class of supernovae, such as SN 2006gy, SN 2006tf and SN 1997cy,  SN 2010jl is the most luminous X-ray SN so far.
\citet{ofek14} reported  
simultaneous \nustar   and \xmm  observations and determined the
temperature of the shock. 
Because of the high temperature of the supernova emission, the hard X-ray sensitivity
of \nustar was crucial to obtain a reliable temperature estimate.
They estimated the shock velocity to be $\sim3000$ $\kms$.
 Given the estimate of the shock velocity and the total luminosity of the SN,
it is possible to estimate the density profile
of the CSM if the forward shock
wave is radiative. With this assumption the presupernova star lost $\sim 3-10\Msun$
in the decades prior to the explosion \citep{zhang12,ofek14,fec+14}.

The H lines in \sn showed a narrow component with an expansion  velocity
$\sim 100$ $\kms$ coming from the CSM; along with
broad wings  which,
at early times, are well fitted by an electron 
scattering profile produced by the
thermal velocities of electrons \citep{fec+14,zhang12}.
Here, the line profiles do not 
reflect the bulk motions of the SN and the high velocity
regions are presumably obscured by the circumstellar gas.
However, over the first 200 days, the broad 
component shifts to the blue by $\sim 700\kms$
\citep{fec+14}.
 \citet{smith12}, \citet{maeda13}, and  \citet{gall14} have explained this  shift to be          
 due to 
the formation of dust in the dense shell resulting from
circumstellar interaction or in the freely 
expanding ejecta.
However, \cite{fec+14} argue that dust 
formation is unlikely and attribute the line
shift to radiative acceleration of circumstellar gas;  
in this case, the broadening of
the lines is due to electron scattering.
Although the situation with the H lines 
is ambiguous, there is clearer evidence for high
velocity motion in the  He~I~$\lambda10830$ line.
\citet{borish14} find a blueshifted shoulder in the $\lambda10830$ line
between 100 and 200 days  that is likely due to 
the ejecta emission up to a velocity of $4000-6000\kms$, in rough agreement with the
velocity deduced from the temperature of X-ray emitting gas at a later time.                


In this paper, we carry out a comprehensive 
analysis of all the  X-ray and radio observations for \sn.
 The radio detection is
being reported for the first time. 
In \S 2, we provide details of observations for \sn. In
\S 3 we present analysis and interpretation of the 
X-ray emission and in \S 4 for the radio data. In \S 5  we discuss
our main results and interpretation in view of 
multiwaveband data. The main conclusions are listed in \S 6.

\section{OBSERVATIONS}
\label{observation}

\subsection{X-ray Observations}

The \swift detection of SN 2010jl allowed us to trigger
 our approved {\em Chandra}  Cycle 11
program and the first observations were made             
on 2010 December 7  and 2010 December 8 for
19 and 21 ks, respectively. The observations were
made using the ACIS-S detector with no grating in the VFAINT
mode. Afterwards, we observed SN 2010jl
on 2011 October 17 (41 ks exposure) and 2012 June 10 (40 ks exposure)
 under Cycle 13 using {\em Chandra}'s
ACIS-S detector.
Our most recent observation was on 2014 June 1 for a
40 ks exposure under Cycle 15.
We also  observed \sn with  \xmm for a 52.2 ks duration
on 2013 November 1.
In addition, we use publicly available archival data from 
HEASARC\footnote{heasarc.gsfc.nasa.gov}. These were 10 ks {\em Chandra}
data on 2010 Nov 22, 12.9 ks {\em XMM-Newton} data
observed on 2012 November 1  and \nustar
data observed for
46 ks on 2013 October 5 \citep{ofek14}, as well as
several {\em Swift}-XRT data sets taken between
2010 November 5 and 2014 December 24. Table \ref{x-ray} gives details of all
the X-ray observations used in this paper.

\begin{deluxetable*}{lccccl}
\tablecaption{Details of X-ray observations for SN 2010jl
\label{x-ray}}
\tablewidth{0pt}
\tablehead{
\colhead{Date of} & \colhead{Mission} & \colhead{Instrument} & \colhead{Obs.} & \colhead{Exposure} &  \colhead{$\Delta t_{exp}$} \\
\colhead{Observation} &\colhead{} & \colhead{} & \colhead{ID} & \colhead{ks}
 & \colhead{days}\tablenotemark{a}}
\startdata
 2010 Nov 05.02 & \emph{Swift} & XRT & 00031858001 & 1.38 &  36.0\\
 2010 Nov 05.08 & \emph{Swift} & XRT & 00031858002 & 1.98&  36.1\\
 2010 Nov 05.67 & \emph{Swift} & XRT & 00031858003 & 6.96&  36.7\\
 2010 Nov 05.88 & \emph{Swift} & XRT & 00031858004 & 4.81&  36.9\\
 2010 Nov 06.08 & \emph{Swift} & XRT & 00031858005 & 1.97&  37.1\\
 2010 Nov 07.02 & \emph{Swift} & XRT & 00031858006 & 2.33&  38.0\\
 2010 Nov 08.16 & \emph{Swift} & XRT & 00031858007 & 2.80&  39.2\\  
 2010 Nov 09.02 & \emph{Swift} & XRT & 00031858010 & 1.68&  40.0\\
 2010 Nov 09.10 & \emph{Swift} & XRT & 00031858008 & 0.18&  40.1\\
 2010 Nov 09.10 & \emph{Swift} & XRT & 00031858009  & 0.58&   40.1\\
 2010 Nov 11.06 & \emph{Swift} & XRT & 00031858011 & 2.30&  42.q\\
 2010 Nov 12.03 & \emph{Swift} & XRT & 00031858012 & 0.56&  43.0\\
 2010 Nov 12.03 & \emph{Swift} & XRT & 00031858014 & 11.13&  43.0\\
 2010 Nov 12.16 & \emph{Swift} & XRT & 00031858013 & 2.09&  43.2\\
 2010 Nov 13.78 & \emph{Swift} & XRT & 00031858015 & 2.22&  44.8\\
 2010 Nov 14.11 & \emph{Swift} & XRT & 00031858016 & 2.41&  45.1\\
 2010 Nov 15.58 & \emph{Swift} & XRT & 00031858017 & 1.83&  46.6\\
 2010 Nov 16.71 & \emph{Swift} & XRT & 00031858018 & 2.29&  47.7\\
 2010 Nov 17.45 & \emph{Swift} & XRT & 00031858019 & 2.13&  48.5\\
 2010 Nov 20.07 & \emph{Swift} & XRT & 00031858020 & 2.11&  51.1\\
2010 Nov 22.03 & \emph{Chandra} & ACIS-S & 11237 & 10.05 &  53.0\\
 2010 Nov 23.01 & \emph{Swift} & XRT & 00031858021 & 2.41&  54.0\\
 2010 Nov 26.16 & \emph{Swift} & XRT & 00031858022 & 2.41&  57.2\\
 2010 Nov 29.69 & \emph{Swift} & XRT & 00031858023 & 2.27&  60.7\\
 2010 Dec 02.13 & \emph{Swift} & XRT & 00031858024 & 2.09&  63.1\\
 2010 Dec 05.67 & \emph{Swift} & XRT & 00031858025 & 2.42&  66.7\\
2010 Dec 07.18 & \emph{Chandra} & ACIS-S & 11122 & 19.05 &  68.2\\
2010 Dec 08.03 & \emph{Chandra} & ACIS-S & 13199 & 21.05 &  69.0\\
2011 Apr 24.56 & \emph{Swift} & XRT & 00031858026 & 7.89&  206.6\\
2011 Apr 28.04 & \emph{Swift} & XRT & 00031858027 & 2.28&  210.0\\
2011 Oct 17.85 & \emph{Chandra} & ACIS-S & 13781 & 41.04&  382.9\\
2012 Jun 10.67 & \emph{Chandra} & ACIS-S & 13782 & 40.07&  619.7\\
2012 Oct 05.98 &  \emph{NuSTAR} &  FPMA & 40002092001 & 46.11& 737.0\\
2012 Oct 05.98 &  \emph{NuSTAR} &  FPMB & 40002092001 & 46.07& 737.0\\
2012 Oct 07.02 & \emph{Swift} & XRT & 00080420001 & 2.65&  738.0\\
2012 Oct 21.31 & \emph{Swift} & XRT & 00032585001 & 8.08&  752.3\\
2012 Nov 01.63 & \emph{XMM-Newton} &EPIC-PN   & 0700381901 & 4.04& 763.6\\
2012 Nov 01.63 & \emph{XMM-Newton} &EPIC-MOS1   & 0700381901 & 10.11& 763.6\\
2012 Nov 01.63 & \emph{XMM-Newton} &EPIC-MOS2   & 0700381901 & 9.73& 763.6\\
2013 Jan 21.10 & \emph{Swift} & XRT & 00032585002 & 8.19&  844.1\\
2013 Feb 10.00 & \emph{Swift} & XRT & 00032585003 & 4.78&  864.0\\
2013 Feb 20.61 & \emph{Swift} & XRT & 00032585004 & 5.90&  874.6\\
2013 Mar 04.49 & \emph{Swift} & XRT & 00046690001 & 0.92&  886.5\\
2013 Mar 29.27 & \emph{Swift} & XRT & 00032585005 & 18.63&  911.3\\
2013 May 14.68 & \emph{Swift} & XRT & 00032585006 & 6.89&  957.7\\
2013 May 15.34 & \emph{Swift} & XRT & 00032585007 & 5.47&  958.3\\
2013 May 19.95 & \emph{Swift} & XRT & 00032585008 & 4.82&  963.0\\
2013 May 21.48 & \emph{Swift} & XRT & 00032585009 & 4.76&  964.5\\
2013 Jun 28.00 & \emph{Swift} & XRT & 00032585010 & 8.20&  1002.0\\
2013 Jun 28.34 & \emph{Swift} & XRT & 00032585011 & 6.29&  1002.3\\
2013 Nov 01.67 & \emph{XMM-Newton} &EPIC-PN   & 0724030101 & 52.30& 1128.7\\
2013 Nov 01.67 & \emph{XMM-Newton} &EPIC-MOS1   & 0724030101 & 52.30&  1128.7\\
2013 Nov 01.67 & \emph{XMM-Newton} &EPIC-MOS2   & 0724030101 & 52.30&  1128.7\\
2013 Dec 11.54 & \emph{Swift} & XRT & 00032585012 & 2.07&  1168.5\\
2013 Dec 18.00 & \emph{Swift} & XRT & 00031858013 & 0.95&  1175.0\\
2013 Dec 19.07 & \emph{Swift} & XRT & 00032585014 & 1.21&  1176.1\\
2013 Dec 20.00 & \emph{Swift} & XRT & 00032585015 & 3.20&  1177.0\\
2013 Dec 24.40 & \emph{Swift} & XRT & 00032585016 & 6.68&  1181.4\\
2013 Dec 30.67 & \emph{Swift} & XRT & 00032585017 & 0.59&  1187.7\\
2014 May 11.12 & \emph{Swift} & XRT & 00032585018 & 5.07&  1319.1\\
2014 May 13.04 & \emph{Swift} & XRT & 00032585019 & 1.02&  1321.0\\
2014 May 14.64 & \emph{Swift} & XRT & 00032585020 & 2.24&  1322.6\\
2014 May 16.84 & \emph{Swift} & XRT & 00032585021 & 3.10&  1324.8\\
2014 May 18.78 & \emph{Swift} & XRT & 00032585022 & 0.58&  1326.8\\
2014 May 20.44 & \emph{Swift} & XRT & 00032585023 & 0.32&  1328.4\\
2014 Jun 01.25 & \emph{Chandra} & ACIS-S & 15869 & 40.06 &  1340.3\\
2014 Jun 27.90 & \emph{Swift} & XRT & 00046690002 & 0.92 & 1366.9\\
2014 Nov 30.76 & \emph{Swift} & XRT & 00032585024 & 1.27 & 1522.7\\
2014 Dec 04.22 & \emph{Swift} & XRT & 00032585026 & 1.04 & 1526.2\\
2014 Dec 05.08 & \emph{Swift} & XRT & 00032585027 & 1.51 & 1527.1\\
2014 Dec 08.54 & \emph{Swift} & XRT & 00032585028 & 3.35 & 1530.5\\
2014 Dec 09.01 & \emph{Swift} & XRT & 00032585029 & 2.39 & 1531.0\\
2014 Dec 10.00 & \emph{Swift} & XRT & 00032585030 & 0.35 & 1532.0\\
2014 Dec 18.51 & \emph{Swift} & XRT & 00032585031 & 2.84 & 1540.5\\
2014 Dec 19.52 & \emph{Swift} & XRT & 00032585032 & 1.52 & 1541.5\\
2014 Dec 24.50 & \emph{Swift} & XRT & 00032585033 & 2.92 & 1548.5\\
\enddata
\tablenotetext{a}{Assuming 2010 October 1 to be \sn explosion date.}
\end{deluxetable*}

For the \chandra data analysis, 
we extracted spectra, response and ancillary matrices using 
Chandra Interactive Analysis of Observations software \citep[CIAO;][]{frus06}, using task
{\it specextractor}. The 
CIAO version 4.6 along with CALDB version 4.5.9 was used for this
purpose. To extract the spectra and response matrices for \xmm data,
the Scientific Analysis System (SAS) version 
12.0.1 and its standard commands were used. 
We extracted the spectra from \nustar data using The  NuSTAR  Data  Analysis  Software (NUSTARDAS) version 1.3.1.
The task {\it nupipeline} was used to generate level 2 products and
{\it nuproducts} was used to generate level 3 spectra and
matrices. The \swift spectra and response matrices were extracted using 
online XRT products building
pipeline\footnote{\url{http://www.swift.ac.uk/user\_objects/}} 
\citep{evan09,goad07}. The 
HEAsoft\footnote{\url{http://heasarc.gsfc.nasa.gov/docs/software/lheasoft/}}
package xspec version 12.1 \citep{arnaud96}
was used to carry out the spectral analysis. 

\subsection{Radio Observations}
The radio observations of SN 2010jl were carried out using the Expanded Very Large Array (EVLA) telescope,
later renamed to Karl G. Jansky Very Large Array (JVLA), starting from   
2010 November 6 until 
  2013 August 10. The observations were carried out at 33 GHz (Ka band), 22 GHz (K band), 
8.5 GHz (X band) and 5 GHz (C band) frequency bands 
for 30 minute to 1 hour durations. 
Each observation consisted of the flux calibrator 3C286  and a 
phase calibrator. The phase calibrator was  J1007+1356 in most cases,  and
 J0954+1743 in a few cases. The bandwidths used in the EVLA data and JVLA data were
  256 MHz and
  2048 MHz, respectively. In some of the EVLA observations, 
each 128 MHz subband was tuned to 4.5 and 7.5 GHz bands in order to  
estimate the flux density at the above two frequencies. The 
data were analyzed using 
the Common Astronomy Software Applications \citep[CASA;][]{casa07}. The VLA 
Calibration pipeline\footnote{\url{https://science.nrao.edu/facilities/vla/data-processing/pipeline}} was used for flagging and calibration purposes.  However,
in several cases, extra flagging was needed. In those cases, flagging and  calibration was done manually. The images were made with CASA 
task `clean' in which  ``briggs'' weighting with
 a robustness parameter of 0.5 was used. 
For the 2 GB bandwidth data, ``mfs'' spectral gridding mode 
with two Taylor coefficients was used to model the sky frequency dependence.
The observational details are presented in Table \ref{tab:radio}.

\begin{deluxetable*}{llcccc}
\tabletypesize{\scriptsize}
\tablecaption{0.2--10~keV X-ray fluxes of  UGC 5189A in \chandra~ observations
\label{tab:flux}}
\tablewidth{0pt}
\tablehead{
\colhead{Date of} & \colhead{$\Delta t_{exp}$} & \colhead{Count} & \colhead{Abs. Flux} & 
\colhead{Unabs. Flux} & \colhead{Unabs. Luminosity} \\
\colhead{Observation} &\colhead{days\tablenotemark{a}} & \colhead{Rate} & \colhead{erg cm$^{-2}$ s$^{-1}$} & 
\colhead{erg cm$^{-2}$ s$^{-1}$} & \colhead{erg s$^{-1}$}
}
\startdata
2010 Nov 22.03 & 53 & $(1.50\pm0.39)\times10^{-3}$ & $(2.93\pm0.66)\times 10^{-14}$ & $(3.23\pm0.73)\times 10^{-14}$ & $(9.27\pm2.09)\times10^{39}$\\ 
2010 Dec 07.18--8.03 & 68.2--69.0& $(0.87\pm0.15)\times10^{-3}$ & $(1.61\pm0.25)\times 10^{-14}$ & $(1.77\pm0.28)\times 10^{-14}$ & $(5.10\pm0.79)\times10^{39}$\\
2011 Oct 17.85 &382.9 &  $(1.07\pm0.17)\times10^{-3}$ & $(1.99\pm0.28)\times 10^{-14}$ & $(2.19\pm0.31)\times 10^{-14}$ & $(6.29\pm0.90)\times10^{39}$\\
2012 Jun 10.67 & 619.7 & $(1.37\pm0.19)\times10^{-3}$ & $(2.63\pm0.33)\times 10^{-14}$ & $(2.90\pm0.37)\times 10^{-14}$ & $(8.34\pm1.05)\times10^{39}$\\
2014 Jun 01.25 & 1340.3 & $(0.92\pm0.15)\times10^{-3}$ & $(1.69\pm0.26)\times 10^{-14}$ & $(1.86\pm0.28)\times 10^{-14}$ & $(5.34\pm0.81)\times10^{39}$\\
Joint fit  & $\cdots$  & $\cdots$ & $(2.02\pm0.14)\times10^{-14}$ & 
$(2.22\pm0.16)\times10^{-14}$ &
$(6.13\pm0.44)\times10^{39}$
\enddata
\tablenotetext{a}{Assuming 2010 October 1 to be \sn explosion date.}
\tablecomments{ The fluxes are derived as detailed in \S \ref{sec:other}.}
\end{deluxetable*}

\begin{deluxetable*}{llcccc}
\tabletypesize{\scriptsize}
\tablecaption{0.2--10~keV fluxes of  6 sources (Fig. \ref{fig:ds9}, right panel)
within $21''$ radius of the \sn position in  \chandra observations                    
\label{tab:flux2}}
\tablewidth{0pt}
\tablehead{
\colhead{Date of}  & \colhead{$\Delta t_{exp}$} & \colhead{Count} & \colhead{Abs. Flux} & 
\colhead{Unabs. Flux} & \colhead{Unabs. Luminosity} \\
\colhead{Observation} &\colhead{days\tablenotemark{a}}&\colhead{Rate} & \colhead{erg cm$^{-2}$ s$^{-1}$} & 
\colhead{erg cm$^{-2}$ s$^{-1}$} & \colhead{erg s$^{-1}$}
}
\startdata
2010 Nov 22.03 & 53& $(2.28\pm0.58)\times10^{-3}$ & $(2.17\pm0.45)\times 10^{-14}$ & $(3.19\pm0.66)\times 10^{-14}$ & $(8.81\pm1.82)\times10^{39}$\\ 
2010 Dec 07.18--8.03& 68.2--69.0 & $(1.68\pm0.26)\times10^{-3}$ & $(1.51\pm0.19)\times 10^{-14}$ & $(2.21\pm0.28)\times 10^{-14}$ & $(6.10\pm0.76)\times10^{39}$\\
2011 Oct 17.85 &382.9& $(3.53\pm0.34)\times10^{-3}$ & $(3.17\pm0.27)\times 10^{-14}$ & $(4.64\pm0.40)\times 10^{-14}$ & $(12.84\pm1.10)\times10^{39}$\\
2012 Jun 10.67 & 619.7 & $(2.23\pm0.30)\times10^{-3}$ & $(2.27\pm0.24)\times 10^{-14}$ & $(3.33\pm0.35)\times 10^{-14}$ & $(9.21\pm0.95)\times10^{39}$\\
2014 Jun 01.25 & 1340.3& $(2.82\pm0.28)\times10^{-3}$ & $(2.56\pm0.24)\times 10^{-14}$ & $(3.75\pm0.36)\times 10^{-14}$ & $(10.37\pm0.99)\times10^{39}$\\
Joint fit  & $\cdots$  &$\cdots$ & $(2.22\pm0.11)\times10^{-14}$ & $(3.30\pm0.17)\times10^{-14}$ &
$(9.11\pm0.47)\times10^{39}$
\enddata
\tablenotetext{a}{Assuming 2010 October 1 to be \sn explosion date.}
\tablecomments{ The fluxes are derived as detailed in \S \ref{sec:other}.}
\end{deluxetable*}

\section{X-RAY ANALYSIS AND INTERPRETATION}
\label{sec:xray}

\subsection{Analysis of the Contaminating Sources}
\label{sec:other}

 \sn host galaxy belongs to
UGC 5189 group of galaxies, which has a size of $1.7'$ 
 centered at
$\alpha=09^h42^m54.^s72$,  $\delta=+09^{\circ}29'01.''4$ 
(J2000).
A NASA Extragalactic Database\footnote{\url{http://ned.ipac.caltech.edu/}} 
(NED) search shows that 
there are  three sources within $3''$  of the \sn position. 
These are   UGC 05189 NED01 (UGC 5189A, $\alpha=09^h42^m53^s.434$, 
$\delta=+09^{\circ}29'41.''87$ (J2000)),
MCG +02-25-021 GROUP 
($\alpha=09^h42^m53^s$,  $\delta=+09^{\circ}29.'7$ (J2000)) and 
SDSS J094253.47+092943.5 ($\alpha=09^h42^m53^s.47$, 
$\delta=+09^{\circ}29'43.''51$ (J2000)) at  distances of  $1.44''$, 
$2.34''$ and $2.46''$ away from the SN, respectively.
The SDSS J094253.47+092943.5 and  UGC 05189A sources have been identified as galaxies, whereas MCG +02-25-021 GROUP is  a group of galaxies.
Since there are no X-ray archival data at
the \sn field of view (FoV), we could not ascertain whether these three sources were
X-ray emitters or not. For this reason we started our analysis with the \chandra
data. Since \chandra has excellent spatial resolution, and can separate
out the nearby sources.

\begin{figure*}
\begin{center}
\includegraphics[angle=0,width=0.47\textwidth]{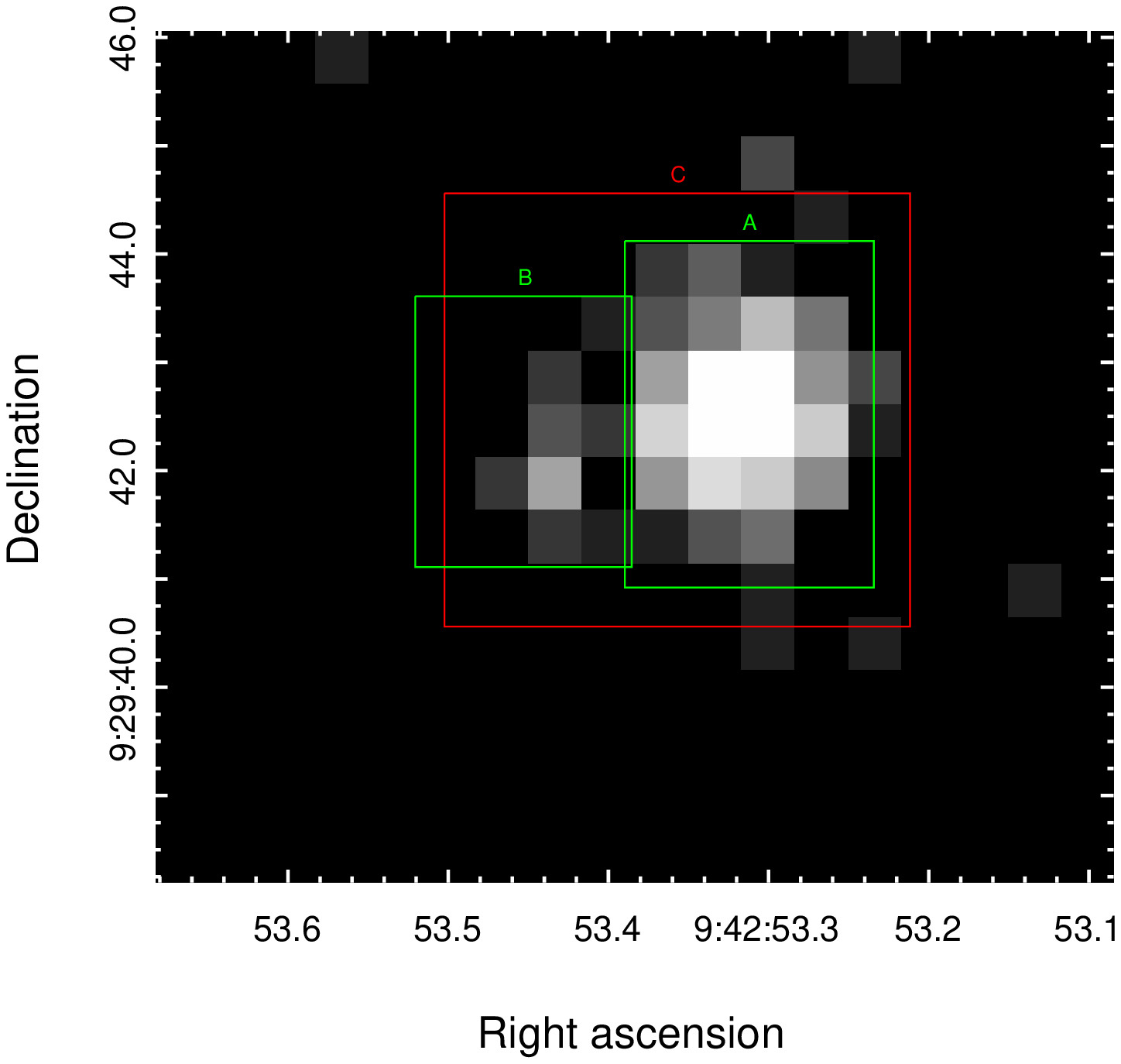}
\includegraphics[angle=0,width=0.52\textwidth]{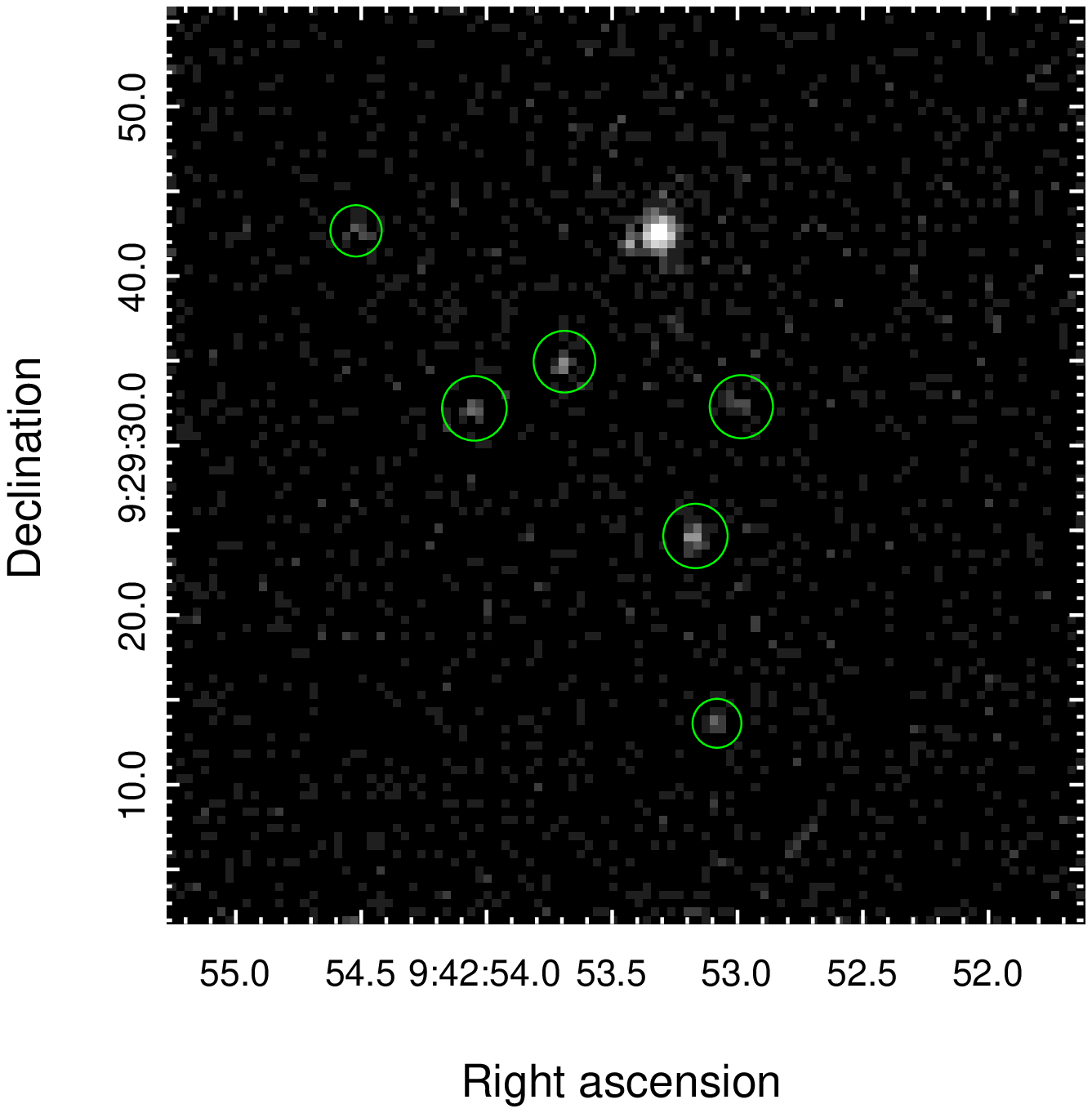}
\caption{{\it Left Panel:} SN 2010jl field of view (FoV) from Chandra 
observations. 
The left green box ($2.0''\times2.5''$) marked `B'
is used to extract the spectrum for UGC 5189A. The right green box ($2.3''\times3.2''$)
marked `A'
has been used to extract the SN 2010jl spectrum. The largest red box 
($4.3''\times4.0''$) marked `C' includes
both SN 2010jl and the UGC 5189A.
{\it Right Panel:} Field of view from Chandra observations. Here \sn is 
the brightest source at the top part of the image center. The green circles are the 6 nearby sources within a $21\arcsec$ radius centered at the \sn
position, excluding  UGC 5189A.}
\label{fig:ds9}
\end{center}
\end{figure*}

The \sn FoV in \chandra observations at various epochs 
show  UGC 5189A to be an X-ray emitter.   However,  no X-ray emission in seen from 
MCG +02-25-021 GROUP or SDSS J094253.47+092943.5.
Thus we need to make sure that the UGC 5189A
 does not contaminate 
 the  \sn flux in \chandra data. We define three boxes in the \chandra
FoV as shown in the left panel of Fig. \ref{fig:ds9}. Box A of
size $2.3''\times3.2''$ covers \snb, while box B of size 
 $2.0''\times2.5''$ covers UGC 5189A.
We also extract a $4.3''\times4.0''$ box covering both \sn and 
UGC 5189A centered at (J2000)
$\alpha=09^h42^m53.^s361$ $\delta=+09^{\circ}29'41.''55$  
(Box C). The background region is chosen 
in a source free area with 
a $9.0''\times9.0''$ box.

In order to estimate the contamination from \ugc, we start our analysis with the \chandra observations on
2010 December 7 and 8. We extract the spectra of the SN alone from Box A, 
and of UGC 5189A
from box B.  We also extract the combined spectrum  from Box C. 
The spectra are grouped into 15 channels for Boxes A and C and $\chi^2$-statistics
is used. 
However, due to a small number of counts in the Box B, the spectrum was binned
into 5 channels per bin and C-statistics were used to fit the data.
 We fit the UGC 5189A spectrum with an absorbed power law model, whereas,
we use the Astrophysical Plasma Emission Code \citep[apec; ][]{smith01}
 to fit the spectrum for \snb.
The apec gives a fit to an emission spectrum from collisionally-ionized diffuse gas.
The parameters of this model are the plasma temperature, 
metal abundances and redshift.
In the spectrum from Box C, 
we fit the joint absorbed thermal plasma and absorbed power law 
spectra. Here we fix the respective parameters to the best
fit values obtained from the spectral fits of Box A and
Box B; however, we let the normalizations vary.  We estimate the 0.2--10 keV
fluxes in Boxes A, B, and C to be 
$(6.58\pm0.38)\times10^{-13}$ erg cm$^{-2}$ s$^{-1}$,
$(1.50\pm0.23)\times10^{-14}$ erg cm$^{-2}$ s$^{-1}$
and $(6.93\pm0.32)\times10^{-13}$ erg cm$^{-2}$ s$^{-1}$, respectively.
The flux in  Box C matches  the total 
flux from Boxes A and  B within the error bars. In addition, the UGC 5189A flux is 40 times weaker than the SN flux.
Thus the contamination to the SN flux due to UGC 5189A is insignificant in
the \chandra data.

Now, we  carry out a joint fit to the
spectra from UGC 5189A (Box B) at all five \chandra 
epochs of observation with an absorbed power law model. 
We assume that the absorption column density and power 
law index do not change
at various epochs and enforce these parameters  to be the
same at all epochs by linking them in the fits.
However, we let normalizations vary independently 
to account for variable X-ray emission.
Since the counts are few, we  use C-statistics to fit the
data.
The model is best fitted by 
an absorbed power law with an absorption column density of
$N_H=(1.82^{+2.00}_{-1.63})\times10^{21}$ cm$^{-2}$
and a photon power law index  $\Gamma=1.16^{+0.43}_{-0.40}$.
This type of photon index is consistent if the X-ray emission in the galaxy is mainly 
from a collection
of X-ray binaries (XRBs) or from a combination of XRBs and diffuse gas.
 The best fit C-statistic is 22.94 for 30 degrees of
freedom. 
The absorption column density is much higher than the Galactic 
absorption which is
$N_H(\rm Galactic) =3\times10^{20}$ cm$^{-2}$. The remaining column
density $N_H(\rm Host)=1.52\times10^{21}$ cm$^{-2}$ must be
coming from the host galaxy UGC 5189A.
To take care of the low metallicity of the host galaxy, we
refit the data with an absorbed power law with column density $N_H$ to be 
$N_H=N_H({\rm Galactic})+N_H(\rm Host)$.  We use solar metallicity for
$N_H(\rm Galactic)$ and fix it to
$3\times10^{20}$ cm$^{-2}$. We fix the metallicity of $N_H(\rm Host)$ to
be 0.3 solar and treat $N_H(\rm Host)$ to be a free parameter. 
The best fit values are
$N_H(\rm Host)=(4.10^{+5.73}_{-4.00})\times10^{21}$ cm$^{-2}$
and $\Gamma=1.14^{+0.42}_{-0.39}$. The increase in $N_H(\rm Host)$
is due to the lower value of metallicity for the host galaxy.
This simply means that the equivalent hydrogen column density has 
to be $1/0.3$ times larger to account for the
 same absorption of X-rays by metals in a $0.3~  Z_\odot$ 
metallicity environment. Our attempt to fit the UGC 5189A spectra
by fixing the column density to that of the Galactic
value results in
a relatively flat power law index $\Gamma=0.85^{+0.20}_{-0.20}$, which is nonphysical. 
There is an additional evidence of higher neutral HI column density towards UGC 5189 from the Giant Metrewave Radio Telescope (GMRT)
21 cm radio data  obtained in 2013  Nov--Dec \citep{jayaram14}. At a position $\alpha=09^h42^m53^s.434$, 
$\delta=+09^o29'41.''87$ (J2000), they find the HI flux to be 
5.8 mJy $\kms$, 
which translates to a HI column density  of $2.4\times10^{21}$ 
cm$^{-2}$. 

\begin{figure}
\centering
\includegraphics[angle=-90,width=0.48\textwidth]{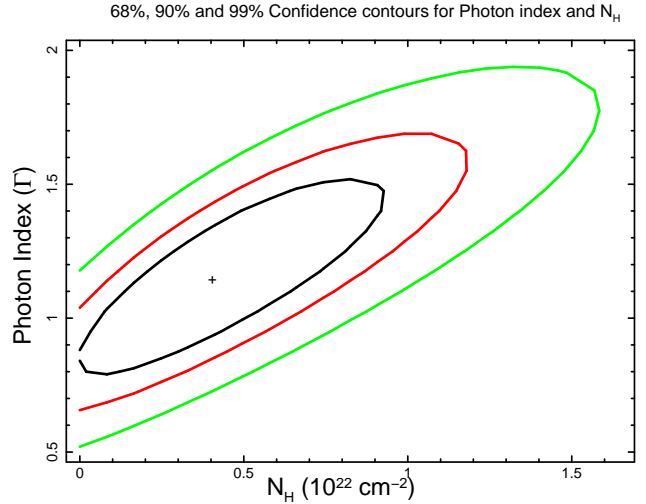}
\caption{The 68\% (black), 90\% (red) and 99\% (green)
confidence contours for best fit
column density ($N_H$) and power law photon index ($\Gamma$) for the UGC 5189A
X-ray spectra obtained from the joint fit of \chandra data at various epochs (see \S \ref{sec:other}). The uncertainty in the column density is large.}
\label{fig:ulx-contour}
\end{figure}

We list the absorbed and unabsorbed fluxes and unabsorbed luminosities
of UGC 5189A in
the 0.2--10~keV range at five \chandra epochs
in Table \ref{tab:flux}. 
The flux 
 varies at most by a factor of 1.8 in these epochs, so we 
  refit the data 
assuming constant flux at all the epochs. 
This also gives a  reasonable fit with
a best fit C-statistic of 31.4 for 36 degrees of freedom.
Here the best fit values are 
$N_H(\rm Host)=(4.07^{+5.71}_{-4.00})\times10^{21}$ cm$^{-2}$
and $\Gamma=1.15^{+0.42}_{-0.39}$. The 0.2--10
keV absorbed (unabsorbed) flux of UGC 5189A is
$(2.02\pm0.14)\times10^{-14}$ erg cm$^{-2}$ s$^{-1}$
($(2.22\pm0.16)\times10^{-14}$ erg cm$^{-2}$ s$^{-1}$), which translates
to an unabsorbed luminosity  
of $(6.13\pm0.44)\times10^{39}$ erg s$^{-1}$ (Table \ref{tab:flux}).
We also plot the contour levels of the
best fit column density and photon index in Figure \ref{fig:ulx-contour}.
We note that there is a large uncertainty in the absorption column density.

The spatial resolution of the \swift, \xmm and \nustar observations are
not as good as that of \chandra, so we have to take care of contamination 
from more distant sources while analyzing these data.
We looked for the contaminating sources within a $60''$  region centered at \snb. 
As shown in the \chandra FOV (right panel of Fig. \ref{fig:ds9}), 
there are 6 sources within $21''$
of the SN position,  in addition to UGC 5189A. There are no additional         
sources between $21''$ and $60''$ radius centered at the SN.
We extracted spectra in the 0.2--10~keV range from \chandra data at each epoch for 
these 6 sources and carried out a joint fit, assuming the flux did not change
at various \chandra epochs.
Their spectra are best fit with a column density of
$N_H(\rm Host)=(4.72^{+3.82}_{-3.06})\times 10^{21}$ cm$^{-2}$
and a power law index of $\Gamma=2.05^{+0.43}_{-0.37}$
(reduced $\chi^2=1.07$). We note that the absorption column density is similar to
that obtained for UGC 5189A,  re-confirming  that
the host galaxy UGC 5189  contributes  a significant amount of 
X-ray absorbing  column.
The 0.2--10~keV absorbed (unabsorbed) flux  of these sources combined is
$(2.22\pm0.11)\times10^{-14}$ erg cm$^{-2}$ s$^{-1}$
($(3.30\pm0.17)\times10^{-14}$ erg cm$^{-2}$ s$^{-1}$), which translates
to an unabsorbed luminosity
of $(9.11\pm0.47)\times10^{39}$ erg s$^{-1}$. We also attempted to
carry out fits where we let normalizations at each epoch vary. The
absorbed flux changed at most by a factor of 2, the reduced $\chi^2=0.84$
improved significantly. In Table \ref{tab:flux2} we give the $0.2-10$~keV
fluxes of these sources.
 
 \subsection{Analysis of SN 2010jl}
\label{sec:sn}

 \citet{chandra12a}  analyzed \chandra data from 2010 December and
2011 October observations. In both cases, their best fit temperature 
values always hit
the hard upper limit of  the models  in XSPEC.
The power law models  were discarded since they gave an unphysically hard spectrum.         
\citet{chandra12a} preferred a thermal model, noting  that the plasma giving rise to 
the X-ray emission is sufficiently hot  that 
\chandra  is not sensitive to the high plasma temperature.   However, we now have the advantage of having \sn observations with 
\nustar which has   sensitivity in the range 3--80~keV. 
Because of the  above complications, we             
 use \nustar data to determine the shock temperature and then use the same 
temperature for the analysis of      
the rest of the data.

\subsubsection{NuSTAR data}
\label{sec:nustar}

\nustar observed  \sn
on 2012 October 5 \citep{ofek14}. \sn was also observed with 
\xmm on 2012 November 1. To get  spectral coverage in a wider energy range, we
 carried out a joint fit to both \nustar and \xmm  spectra. 
We used the best fit parameters  of the
2012 June 1 \chandra data  for UGC 5189A and the other 
6 contaminating sources to subtract out their contamination to SN
flux (see \S \ref{sec:other}).
For the \xmm observations, we used spectra from MOS1, MOS2 and 
PN CCD arrays, whereas for the \nustar observations, we
used spectra from both FPMA as well as FPMB focal plane modules. 
The total absorption column density in the models are
$N_H$, where $N_H=N_H(\rm Galactic)+N_H(\rm Host)+N_H(\rm CSM)$. 
Here $N_H(\rm Galactic)$ and $N_H(\rm Host)$ are explained in \S \ref{sec:other}
and $N_H(\rm CSM)$ is the column density due to the SN CSM. For $N_H(\rm CSM)$, we
fix the metallicity to be 0.3 solar.
The data are best fit with an
absorbed thermal plasma model. 
The reduced $\chi^2$ is 0.97 for 165 degrees of freedom.
The best fit column density for 
\sn is $N_H(\rm CSM)=(6.67^{+2.47}_{-1.94})\times10^{21}$ cm$^{-2}$
and the plasma temperature is $kT=18.99^{+8.75}_{-4.86}$ keV.
 This temperature is consistent with that found by \cite{ofek14}.

\begin{figure*}
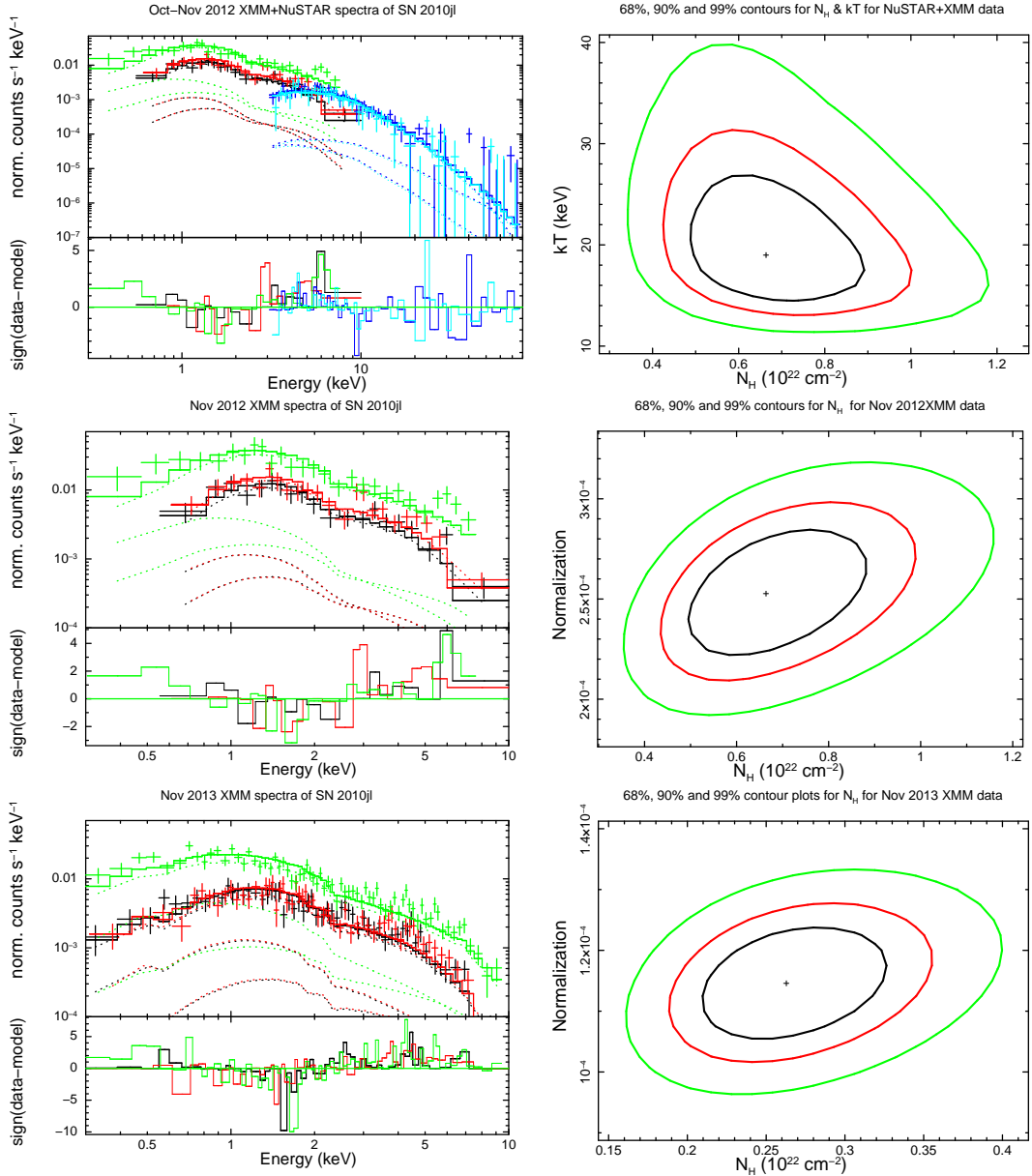

\centering
\includegraphics[angle=-90,width=0.41\textwidth]{f3a.eps}
\includegraphics[angle=-90,width=0.38\textwidth]{f3b.eps}
\includegraphics[angle=-90,width=0.41\textwidth]{f3c.eps}
\includegraphics[angle=-90,width=0.38\textwidth]{f3d.eps}
\includegraphics[angle=-90,width=0.41\textwidth]{f3e.eps}
\includegraphics[angle=-90,width=0.38\textwidth]{f3f.eps}
\caption{\scriptsize{The spectra of \sn and their best fit models as described in \S~\ref{sec:sn}. 
{ \it Upper panel:} The  best fit model
 for the 2012 October~6--November~1 joint \xmm (green: PN, black: MOS1,
 red: MOS2) and
\nustar (blue: FPMA, cyan: FPMB) spectra.  The data are best fit with an absorbed apec model. The right plot shows the 68\% (black), 90\% 
(red) and 99\% (green) confidence contours for the best fit
CSM column density ($N_H$) and the plasma temperature ($kT$). {\it Middle panel:} The
spectrum and column density confidence contours for 
2012 Nov 1.63 \xmm data.
 {\it Lower panel:} The
spectrum and column density confidence contours for 2013 Nov 1.67 
\xmm data. In all the cases, $N_H$ is well constrained.
The colors in the middle and lower panels have the same association as 
explained in the top panel. 
Here the normalized counts s$^{-1}$ kev$^{-1}$ in the 
y-axes in the left plots of the spectra are normalized counts s$^{-1}$ kev$^{-1}$.                
The word normalized indicates that this plot has been divided by the effective area, the value of the EFFAREA keyword, in the response file associated with each spectrum. In all the left side plots,  the residuals are in terms of $\sigma s$ with error bars of size one.} }      
\label{fig:nustar}
\end{figure*}

 Since \citet{chandra12a} claimed the presence of a 6.33~keV Fe         
 $K\alpha$ line in their December 2010 \chandra spectrum, we 
 also attempted to 
 add a Gaussian component around the same energy and refit the spectra. 
 There was no significant change in the quality of the fit. Thus
 Fe K-$\alpha$ line is not significant here.
In Figure \ref{fig:nustar}, we plot the best fit model as well as
the contour diagram of  $N_H (\rm CSM)$ versus $kT$. The 0.2--80
keV absorbed (unabsorbed) flux of the
SN is 
$(5.24\pm0.25)\times10^{-13}$ erg cm$^{-2}$ s$^{-1}$
($(6.01\pm0.28)\times10^{-13}$ erg cm$^{-2}$ s$^{-1}$).
The 0.2--10
keV absorbed (unabsorbed) flux of the
SN  is listed in Table \ref{tab:sn-flux}.

\begin{turnpage}
\begin{deluxetable}{lclllll}[t]
\tabletypesize{\small}
\tablecaption{Best fit models at various epochs
\label{tab:analysis}}
\tablewidth{0pt}
\tablehead{
\colhead{Date of} & \colhead{Instrument} & \colhead{Model}&\colhead{Param-1} & \colhead{Param-2}
& \colhead{Param-3} &\colhead{Reduced}\\
 \colhead{Obsn} &
\colhead{} & \colhead{} & \colhead{$N_H$ (cm$^{-2}$)} &
\colhead{${N_H}_3$ (cm$^{-2}$)} & \colhead{} &\colhead{$\chi^2$}
}
\startdata
2010 Nov 5--20 & \swift & $\sum\limits_{i=1}^2{N_H}_i^**\Gamma_i+N_H^** (kT+
\rm Gauss)$ &
$(17.67^{+16.83}_{-7.91})\times10^{23}$  &$\cdots$ & Gauss$=6.39^{+0.19}_{-0.23}$ & 1.73\tablenotemark{b}\\
\hline
2010 Nov 22.03 & \chandra & $N_H^** kT$ & $(9.59^{+2.60}_{-2.16})\times10^{23}$  
 & $\cdots$ & $\cdots$ & 1.57\tablenotemark{b} \\
\hline
2010 Nov 23--Dec 05 & \swift & $\sum\limits_{i=1}^2{N_H}_i^**\Gamma_i+N_H^** kT$
& $9.59\times10^{23}$ (fixed) & $\cdots$ & $\cdots$ & 1.35\tablenotemark{b} \\
\hline
2010 Dec 7--8 & \chandra & $N_H^** (kT+\rm Gauss)$ & $(9.47^{+0.55}_{-0.52})\times10^{23}$  
 & $\cdots$ & Gauss$=6.33^{+0.07}_{-0.05}$ & 1.90 \\
\hline
2011 Apr 24--28 & \swift & $\sum\limits_{i=1}^2{N_H}_i^**\Gamma_i+N_H^** kT$ & $(5.74^{+2.75}_{-2.00})\times10^{23}$  
 & $\cdots$ & $\cdots$ & 0.92\tablenotemark{b} \\
\hline
2011 Oct 17.85 & \chandra &  $N_H^** kT$ &\tablenotemark{a}$(1.63)\times10^{23}$  & 
 $\cdots$ & $\cdots$ & 2.07\\ 
 2011 Oct 17.85 & \chandra &$N_H^** kT+{N_H}_3^**\Gamma$ 
& $(2.05^{+0.29}_{-0.24})\times10^{23}$ & 
$4\times10^{21}$ (fixed)  & $\Gamma=1.7$ (fixed) & 1.37\\
\bf 2011 Oct 17.85 & \chandra & $N_H^** kT+{N_H}_3^**\Gamma$ &$(2.89^{+0.26}_{-0.22})\times10^{23}$  & 
 $(3.00^{+1.57}_{-1.20})\times10^{21}$  & 
$\Gamma=1.7$ (fixed) & 1.11\\
\hline
2012 Jun 10.67 & \chandra & $N_H^**kT$ &\tablenotemark{a}$(3.74\times10^{22})$  & 
 $\cdots$ & $\cdots$ & 2.71\\ 
2012 Jun 10.67 & \chandra & $N_H^** kT+{N_H}_3^**\Gamma_2$ &$(1.11^{+0.22}_{-0.18})\times10^{23}$  & 
 $0.4\times10^{21}$ (fixed) & 
$\Gamma_2=1.7$(fixed) & 0.93\\
\bf 2012 Jun 10.67 & \chandra & $N_H^** kT+{N_H}_3^**\Gamma_2$ &$(1.09^{+0.25}_{-0.20})\times10^{23}$  & 
 $(3.70^{+1.88}_{-1.53})\times10^{21}$  & 
$\Gamma_2=1.7$ (fixed) & 0.94\\
\hline
2012 Oct 7--21 & \swift & $\sum\limits_{i=1}^2{N_H}_i^**\Gamma_i+N_H^** kT$ & $6.67^\times10^{21}$ (fixed)
& $\cdots$ & $\cdots$ & 2.86\tablenotemark{b} \\
\hline
2012 Oct 5--Nov 1 & \nustar,\xmm &  $\sum\limits_{i=1}^2{N_H}_i^**\Gamma_i+N_H^**kT$ & $(6.67^{+2.47}_{-1.94})\times10^{21}$ & 
 $\cdots$ & $\cdots$& 0.97 \\
\hline
2013 Jan 5--Mar 29 & \swift & $\sum\limits_{i=1}^2{N_H}_i^**\Gamma_i+N_H^** kT$ & $(4.03^{+3.32}_{-2.23})\times10^{21}$ 
& $\cdots$ & $\cdots$ & 1.48\tablenotemark{b} \\
\hline
2013 May 14--Jun 28 & \swift & $\sum\limits_{i=1}^2{N_H}_i^**\Gamma_i+N_H^** kT$ & $(5.58^{+5.75}_{-2.78})\times10^{21}$ 
& $\cdots$ & $\cdots$ & 1.87\tablenotemark{b} \\
\hline
2013 Nov 1.67 & \xmm & 
$\sum\limits_{i=1}^2{N_H}_i^**\Gamma_i+N_H^**kT$ &$(2.64^{+0.69}_{-0.59})\times10^{21}$ &
  $\cdots$ & $\cdots$ & 1.41\\
\hline
2013 Dec 11-30 & \swift & $\sum\limits_{i=1}^2{N_H}_i^**\Gamma_i+N_H^** kT$ & $2.64\times10^{21}$ (fixed) 
& $\cdots$ & $\cdots$ & 1.22\tablenotemark{b} \\
\hline
2014 Jun 1.25 & \chandra & $N_H^**kT$ &$(6.82^{+3.05}_{-2.25})\times10^{21}$  & 
 $\cdots$ & $\cdots$ & 1.24\\ 
\hline
2014 May 11-Jun 27 & \swift & $\sum\limits_{i=1}^2{N_H}_i^**\Gamma_i+N_H^** kT$ & $6.82\times10^{21}$  (fixed) 
 & $\cdots$ & $\cdots$ & 0.65\tablenotemark{b} \\
2014 Nov 30-Dec 24 & \swift & $\sum\limits_{i=1}^2{N_H}_i^**\Gamma_i+N_H^** kT$ & $ (1.41^{+12.90}_{-1.40})\times10^{21}$   
 & $\cdots$ & $\cdots$ & 0.43\tablenotemark{b} \\
\enddata
\tablecomments{Except for the joint \xmm and \nustar spectrum, which gave us the best fit temperature of $kT=18.99^{+8.75}_{-4.86}$ keV, 
everywhere else the temperature has been 
kept fixed to a value of 19 keV. The $kT$ corresponding to all the models in 
the model corresponds to an apec thermal plasma model. The $i=1$ index
in ${N_H}_i$ is for  
UGC 5189A and  $i=2$ for 6 nearby sources within a $21''$ error circle centered at the SN position. The parameters with no suffix correspond to the main SN component
and parameters with suffix 3 correspond to the extra soft component 
present in the 2011 October and 2012 June  data.
We have used ${N_H}_i^*=3\times10^{20}+{N_H}_i$,
${N_H}^*=3\times10^{20}+{N_H}$ and 
${N_H}_3^*=3\times10^{20}+{N_H}_3$. 
This is to account for the contribution from Galactic absorption whose metallicity is fixed to solar. For host and CSM contributions, the metallicity is fixed to 0.3 solar.
 For data with  multiple fits, the
 models in bold are considered to
be the models best representing the respective spectra.}
\tablenotetext{a}{Since  $\chi^2 > 2$,  XSPEC did not calculate the 
error.}
\tablenotetext{b}{C-statistics were used to fit the data. Equivalent $\chi^2$
are mentioned.}
\end{deluxetable}
\end{turnpage}

\clearpage

\begin{figure}
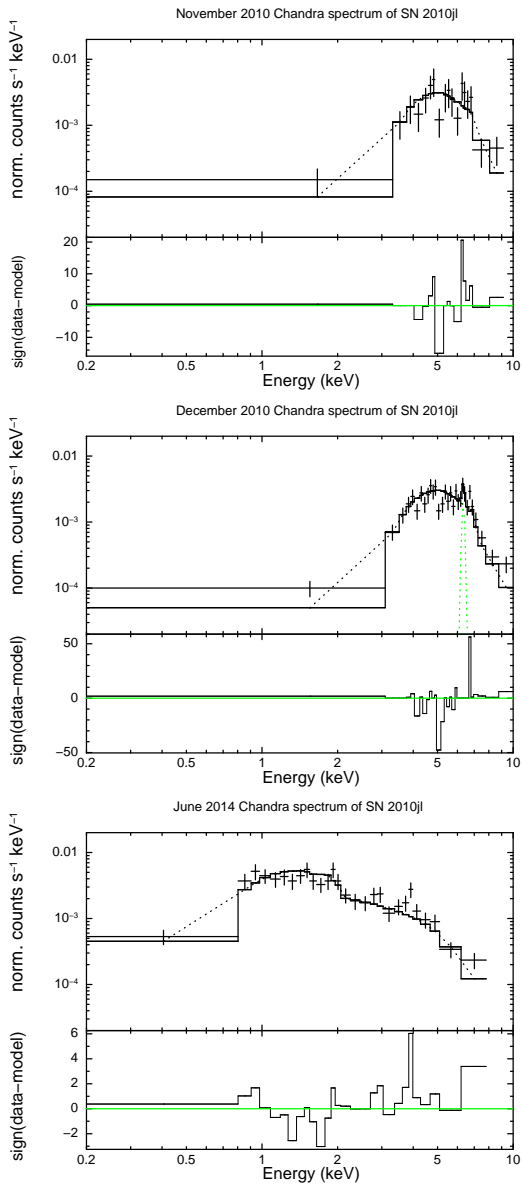

\begin{center}
\includegraphics[angle=-90,width=0.40\textwidth]{f4a.eps}
\includegraphics[angle=-90,width=0.40\textwidth]{f4b.eps}
\includegraphics[angle=-90,width=0.40\textwidth]{f4c.eps}
\caption{The best fit models to \chandra  spectra of \sn at 2010~Nov, 2010~Dec and 2014~June, respectively. 
The best fit models
are explained in \S~\ref{sec:sn} and listed in Table \ref{tab:analysis}.}
\label{fig:sn-dec10}
\end{center}
\end{figure}

\begin{figure*}
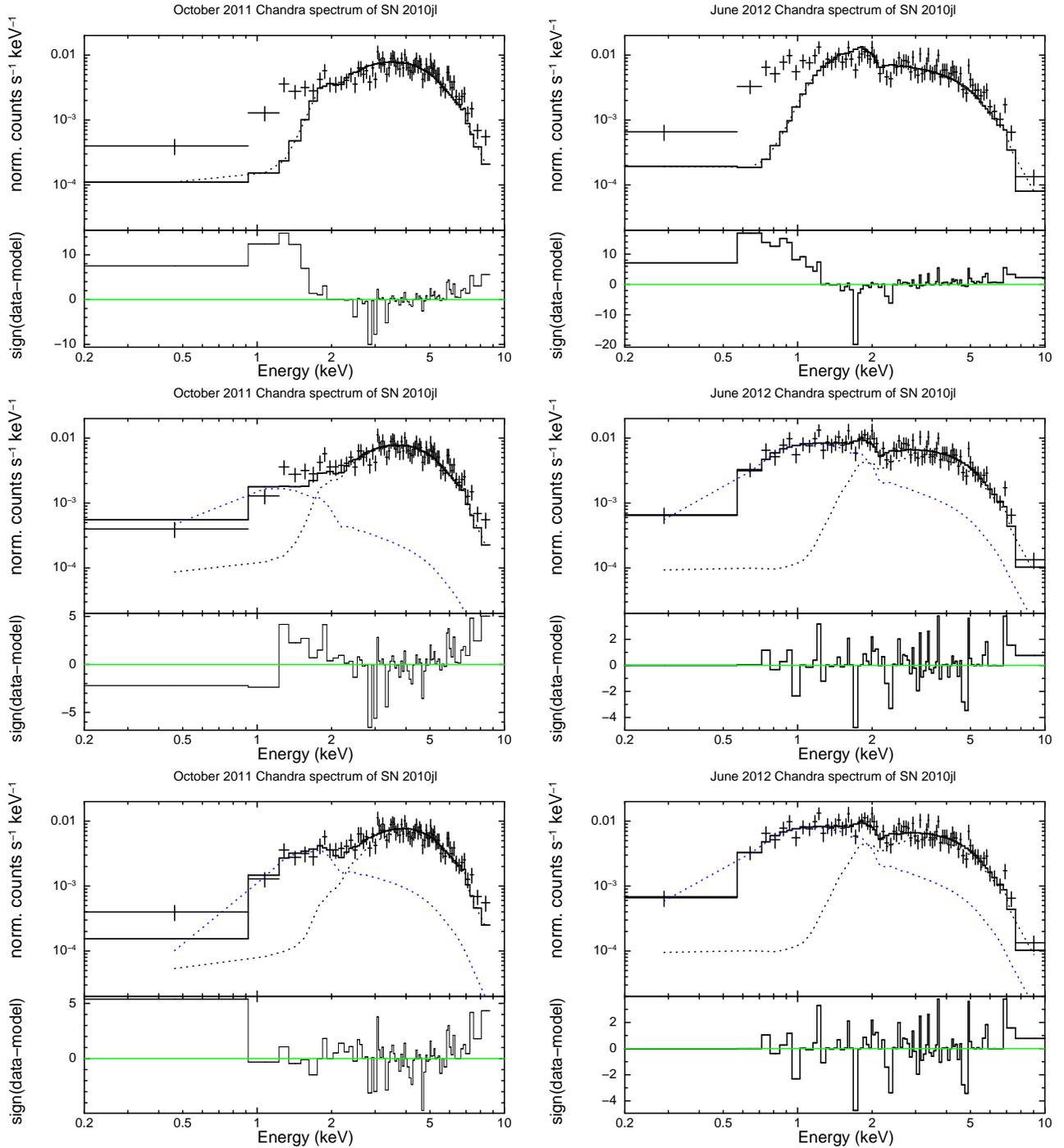

\begin{center}
\includegraphics[angle=-90,width=0.48\textwidth]{f5a.eps}
\includegraphics[angle=-90,width=0.48\textwidth]{f5b.eps}
\includegraphics[angle=-90,width=0.48\textwidth]{f5c.eps}
\includegraphics[angle=-90,width=0.48\textwidth]{f5d.eps}
\includegraphics[angle=-90,width=0.48\textwidth]{f5e.eps}
\includegraphics[angle=-90,width=0.48\textwidth]{f5f.eps}
\caption{\scriptsize{October 2011 (left panels)and June 2012 (right panels)
\chandra spectra of SN 2010jl.
These are the two epochs where the evidence of an extra component is present at the lower energy end of the spectra. {\it Left Panels:} The \chandra best fit spectra for the 2011 October data. The top row is fit with only a thermal plasma model (black dashed line). The middle row includes an additional absorbed power law model (blue dashed line) and fixing the absorption of the extra component to that of host galaxy's absorption, i.e. $4\times10^{21}$ cm$^{-2}$. The lower row  is the same as the middle row, except the column density has been kept as a free parameter.
{\it Right Panels:} Same as left, but for \sn spectra of 2012 June  \chandra data.}}
\label{fig:sn-chandra}
\end{center}
\end{figure*}

\subsubsection{Chandra data}
\label{sec:chandra}

For all data, we fix the temperature to be  $kT=19$ keV (see \S  
\ref{sec:nustar}).  Even though this temperature
 is probably a lower limit
to the temperature for the SN shock at earlier epochs,
it is the best available temperature estimate.
This will probably introduce some errors in the SN flux and the column density estimates. However, we discuss in
\S \ref{sec:results}
that the uncertainties due to the assumption of constant temperature are not significantly large.     

The \chandra data obtained in 2010 November  22 have  very few counts,
 so we group the spectrum
into 5 counts per bin and use C-statistics to fit the data.
We fix all the parameters except the normalization and $N_H$.
For a metallicity of 0.3, the data are best fit with a
column density of $N_H(\rm CSM)=(9.59^{+2.60}_{-2.16})\times10^{23}$ cm$^{-2}$.

The 2010  December 7--8 spectra were grouped in 15 counts per bin and
$\chi^2$-statistics were applied to obtain best fits.  The spectra 
are best fit with a 
column density of 
$N_H(\rm CSM)=(9.47^{+0.55}_{-0.52})\times10^{23}$ cm$^{-2}$.
However, there is an indication of extra emission 
component around 6 keV energy, 
which we also seen by \citet{chandra12a} and was associated with Fe~K-$\alpha$ line. 
In our current fits, the line is best fit with a Gaussian of energy $E_{\rm Gauss}
=6.33^{+0.07}_{-0.05}$ keV and  width  0.19 keV. 
The 0.2--10 keV unabsorbed flux in this line component is 
$(4.59\pm2.53)\times10^{-14}$ erg cm$^{-2}$ s$^{-1}$.

The analysis of 2011
October  data show that the 6.33 keV iron line is not present in the   spectrum, and 
the CSM column density has now decreased to a value of 
$N_H(\rm CSM)=1.63\times10^{23}$ cm$^{-2}$. A decrease of the column density
with time is expected 
as the shock  moves to larger radii. 
However, the apec model does not
 fit the SN 
 spectrum well (reduced $\chi^2=2.07$ for     73
degrees of freedom, also see Fig. \ref{fig:sn-chandra}). There appears 
to be an extra component at the lower energy end of the spectrum. 
We explore three possibilities for this component.
First, the component may be coming from the same region as the harder X-ray
emission, which is most probably the forward shock.
In this case, the column density  for the
soft extra component, $N_H(\rm Soft)$, should be the same as that of the 19 keV component. 
The second possibility is that this component 
is arising from  the reverse shock. In this case the absorption for 
this component should be higher than that of the 19~keV component
as the cool shell will contribute to an  additional absorption. 
In the third case, we let the $N_H$ vary independently.             
The first and second possibilities seem unlikely 
because by fixing the column density to either 
that of the 19~keV component, or declaring 
the  column density associated with the 19~keV component 
to be the lower limit for this extra component (cool 
shell origin of the component), neither
the apec nor 
the power law models give a good fit. The
best fit models result in an  extremely low temperature $<0.1$ keV or
negative power law index, and give a 7 orders of magnitude higher normalization
than the harder component, which is nonphysical. 
When we fit the spectra with a power law index of 
$\Gamma=1.7$ and
fix the  column density for this extra component to
be that of the host galaxy, i.e. $4.1\times10^{21}$ cm$^{-2}$,
the reduced $\chi^2$ improved  
from 2.07 to 1.37.
Allowing  $N_H(\rm Soft)$ to vary freely gives a best fit  
with $N_H(\rm Soft)=3.00^{+1.57}_{-1.20}\times10^{22}$  cm$^{-2}$ and  the
reduced $\chi^2$ improves significantly to 1.11.
In the case of fixing the column density to the host galaxy absorption,
the 0.2--10 keV 
absorbed (unabsorbed) flux of the extra component is
$(3.30\pm0.67)\times10^{-14}$ erg cm$^{-2}$ s$^{-1}$
($(4.14\pm1.84)\times10^{-14}$ erg cm$^{-2}$ s$^{-1}$).
In the case where we let the column density  be a free parameter,
the 0.2--10 keV 
absorbed (unabsorbed) flux of the extra component is
$(1.22\pm0.16)\times10^{-13}$ erg cm$^{-2}$ s$^{-1}$
($(2.01\pm1.27)\times10^{-13}$ erg cm$^{-2}$ s$^{-1}$).

\begin{figure*}
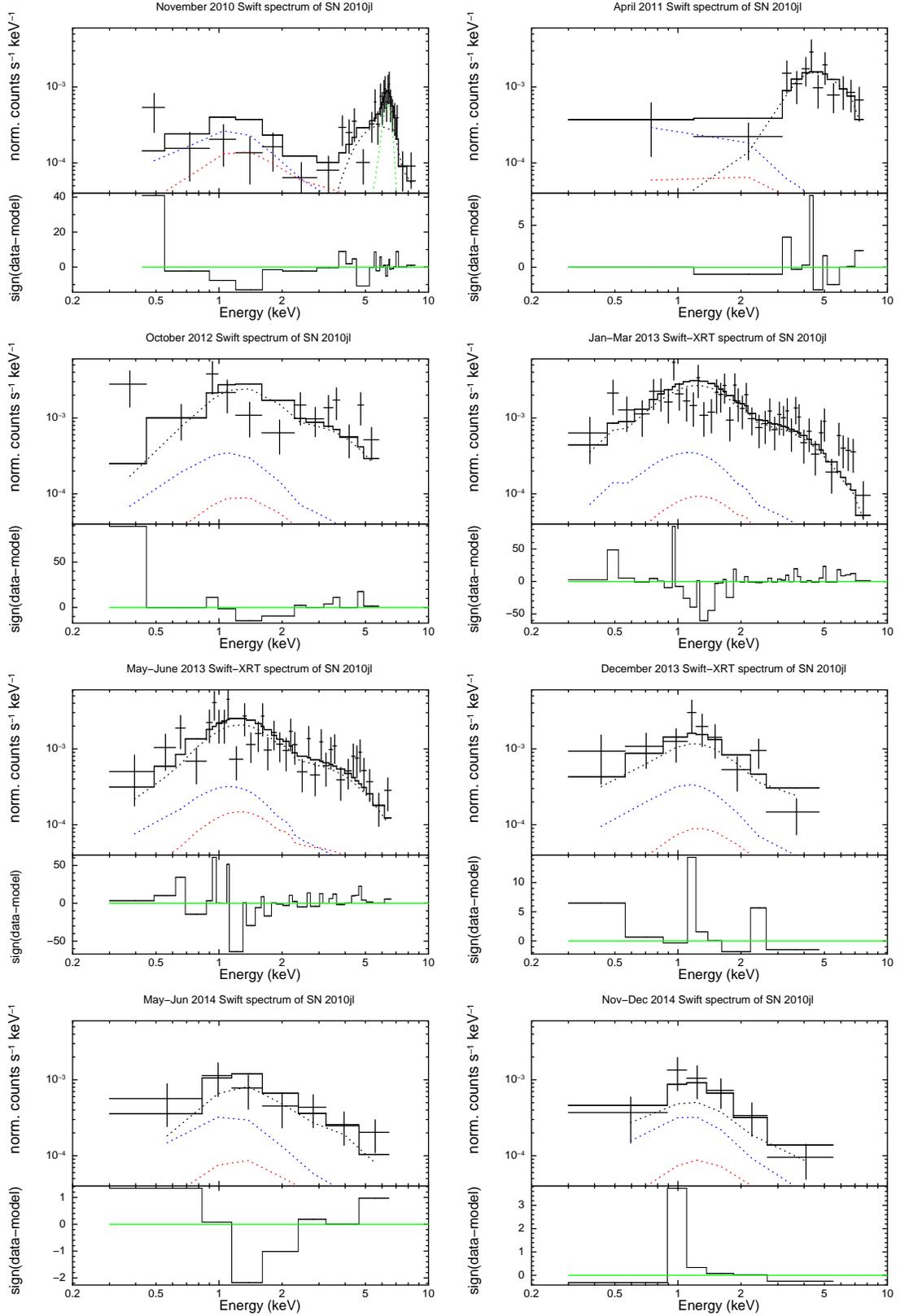

\begin{center}
\includegraphics[angle=-90,width=0.41\textwidth]{f6a.eps}
\includegraphics[angle=-90,width=0.41\textwidth]{f6b.eps}
\includegraphics[angle=-90,width=0.41\textwidth]{f6c.eps}
\includegraphics[angle=-90,width=0.41\textwidth]{f6d.eps}
\includegraphics[angle=-90,width=0.41\textwidth]{f6e.eps}
\includegraphics[angle=-90,width=0.41\textwidth]{f6f.eps}
\includegraphics[angle=-90,width=0.41\textwidth]{f6g.eps}
\includegraphics[angle=-90,width=0.41\textwidth]{f6h.eps}
\caption{The best fit \swift spectra of \sn at various epochs. The best fit models plotted here
are detailed in \S~\ref{sec:sn} and listed Table \ref{tab:analysis}. The black dashed line is the best fit apec thermal plasma model. The red line is the contribution of 
UGC 5189A and the blue line is the contribution from six nearby sources.}
\label{fig:swift}
\end{center}
\end{figure*}

\begin{deluxetable*}{lllccc}
\tabletypesize{\scriptsize}
\tablecaption{SN 2010jl 0.2--10~keV flux at various epochs
\label{tab:sn-flux}}
\tablewidth{0pt}
\tablehead{
\colhead{Date of} & \colhead{Telescope} & \colhead{$\Delta t_{exp}$} &\colhead{Abs. Flux} & 
\colhead{Unabs. Flux} & \colhead{Unabs. Luminosity} \\
\colhead{Observation} &\colhead{} & \colhead{days} &
 \colhead{erg cm$^{-2}$ s$^{-1}$} & 
\colhead{erg cm$^{-2}$ s$^{-1}$} & \colhead{erg s$^{-1}$}
}
\startdata
2010 Nov 5--20 & \swift & $43.55\pm7.53$ & $(4.67\pm0.81)\times 10^{-13}$ & 
$(29.54\pm7.95)\times 10^{-13}$ & $(8.49\pm2.29)\times10^{41}$\\ 
2010 Nov 22.03 & \chandra & 53.03 & $(7.07\pm1.15)\times 10^{-13}$ & 
$(30.56\pm4.95)\times 10^{-13}$ & $(8.78\pm1.42)\times10^{41}$\\ 
2010 Nov 23--Dec 5 & \swift & $60.34\pm6.33$ & $(5.42\pm1.81)\times 10^{-13}$ & 
$(24.33\pm8.13)\times 10^{-13}$ & $(6.99\pm2.33)\times10^{41}$\\ 
2010 Dec 7--8 & \chandra & $68.61\pm0.43$& $(7.03\pm0.52)\times 10^{-13}$ & 
$(29.40\pm2.48)\times 10^{-13}$ & $(8.45\pm0.71)\times10^{41}$\\
2011 Apr 24--28 & \swift & $208.30\pm1.74$ & $(10.70\pm2.39)\times 10^{-13}$ & 
$(34.68\pm7.76)\times 10^{-13}$ & $(9.96\pm2.23)\times10^{41}$\\ 
2011 Oct 17.85  & \chandra &382.85 & $(9.37\pm0.56)\times 10^{-13}$ & $(22.04\pm1.32)\times 10^{-13}$ & $(6.33\pm0.38)\times10^{41}$\\
2012 June 10.67 & \chandra & 619.67& $(5.70\pm0.40)\times 10^{-13}$ & $(10.13\pm0.70)\times 10^{-13}$ & $(2.91\pm0.20)\times10^{41}$\\
2012 Oct 7--21 & \swift & $745.17\pm7.15$ & $(3.48\pm0.81)\times 10^{-13}$ & 
$(4.26\pm1.00)\times 10^{-13}$ & $(1.22\pm0.29)\times10^{41}$\\ 
2012 Oct 5--Nov 1   & \nustar, {\it XMM} &$750.31\pm13.33$ & $(3.64\pm0.17)\times10^{-13}$ & 
$(4.00\pm0.18)\times10^{-13}$ & $(1.15\pm0.06)\times10^{41}$\\
2013 Jan 21--Mar 29 & \swift & $877.69\pm33.59$ & $(3.40\pm0.43)\times 10^{-13}$ & 
$(4.00\pm0.51)\times 10^{-13}$ & $(1.15\pm0.12)\times10^{41}$\\ 
2013 May 14--Jun 28 & \swift & $980.01\pm22.33$ & $(2.73\pm0.39)\times 10^{-13}$ & 
$(3.29\pm0.48)\times 10^{-13}$ & $(0.95\pm0.14)\times10^{41}$\\ 
2013 Nov 1.67 & \xmm & 1128.67&  $(1.88\pm0.07)\times10^{-13}$ &
$(2.15\pm0.08)\times10^{-13}$ & $(0.62\pm0.02)\times10^{41}$\\
2013 Dec 11--30 & \swift 
& $1178.11\pm9.57$ & $(1.50\pm0.49)\times 10^{-13}$ & 
$(1.72\pm0.56)\times 10^{-13}$ & $(0.49\pm0.16)\times10^{41}$\\ 
2014 Jun 1.25 & \chandra &1340.25 & $(1.53\pm0.13)\times 10^{-13}$ & $(1.82\pm0.15)\times 10^{-13}$ & $(0.52\pm0.04)\times10^{41}$\\
2014 May 11--Jun 27 & \swift & $1343.01\pm23.89$ & $(1.22\pm0.47)\times 10^{-13}$ & 
$(1.49\pm0.58)\times 10^{-13}$ & $(0.43\pm0.17)\times10^{41}$\\ 
2014 Nov 30--Dec 24 & \swift & $1535.60\pm12.90$ & $(0.64\pm0.04)\times 10^{-13}$ & 
$(0.72\pm0.44)\times 10^{-13}$ & $(0.21\pm0.13)\times10^{41}$\\ 
\enddata
\tablecomments{Here the fluxes and luminosities are in the 0.2--10 keV
range. The fluxes are derived as detailed in \S \ref{sec:sn}.
The SN explosion date is assumed to be 2010 October 1.}
\end{deluxetable*}

Now we carry out an analysis  of the 2012 June   \chandra spectrum.
Here the column density is best fit  with a value of
$N_H(\rm CSM)=3.74\times10^{22}$ cm$^{-2}$. The 
  low temperature feature seen in the 2011 October 
data is  prominent here as well
and the reduced $\chi^2$ for the absorbed thermal plasma 
is quite large ($\chi^2=2.71$ for     80
degrees of freedom).
 For this additional low temperature component, we explore the  same three possibilities   as discussed in the above paragraph for the
 case of 2011 October  data.
In the first case, when $N_H(\rm Soft)$ is fixed to a value  the same as that for the
19~keV component, i.e. $N_H(\rm CSM)=3.74\times10^{22}$ cm$^{-2}$, the best temperature goes
to very low values $kT\ll 0.1$ keV. In the case of powerlaw fits,
 the power law index becomes negative, 
and the normalization becomes unphysically high.               
When we fit the spectra by
fixing the  column density for the extra component to
be the same as that of host galaxy, the reduced $\chi^2$ improved significantly, from 2.71 to 0.94. 
However, if we let $N_H(\rm Soft)$ vary freely, it did not change significantly from that of host galaxy column density value, and the
reduced $\chi^2$  did not improve. 
When we kept the column density fixed to the host galaxy value, 
the 0.2--10~keV 
absorbed (unabsorbed) flux of the extra component was
$(1.58\pm0.67)\times10^{-13}$ erg cm$^{-2}$ s$^{-1}$
($(1.99\pm0.17)\times10^{-13}$ erg cm$^{-2}$ s$^{-1}$).
If we let the column density vary, the 0.2--10 keV absorbed (unabsorbed)
flux of the component is 
$(1.55\pm0.14)\times10^{-13}$ erg cm$^{-2}$ s$^{-1}$
($(1.93\pm0.17)\times10^{-13}$ erg cm$^{-2}$ s$^{-1}$).

If we assume that the column density of the extra component
is fixed to $N_H(\rm Host)$
in both the 2011 October  and 2012 June  data, then the component is much stronger
at the later epoch.
Surprisingly this component is not present in the 2014 June  \chandra
data. If we let the column density  be a free parameter, then the 
flux is roughly constant. We do not understand a process 
that can increase the flux by almost a factor 5  between 2011 October  and 
2012 June  and 
then vanish completely. Thus we consider the varying column 
density model to be the more realistic model for this extra component. 
As per our fits, the component 
appeared around 2011 October, then became optically thin (reduced 
absorption consistent with the host galaxy absorption), and 
finally disappeared.
The  fits to 2011 October and 2012 June data are plotted in Fig. \ref{fig:sn-chandra}.

In the 2014  June  \chandra data
the SN component is  fitted with the
absorbed apec model and the
best fit column density is 
$N_H(\rm CSM)=(6.82^{+3.05}_{-2.25})\times10^{21}$ cm$^{-2}$.
Initially we let  the temperature be a free parameter, to check if the
forward shock  cooled down significantly or if the  reverse shock
too has started to contribute towards X-ray emission.  But the best fit temperature again hit
the model upper limit of 80~keV,  so we fixed it
to  $kT=19$ keV as in previous datasets.
The best fit absorbed apec model gives a reduced $\chi^2=1.15$.
The   extra component seen in the
2011 October  and 2012 June  data is no longer detectable.
The best fit models are listed in Table \ref{tab:analysis}, and
the best fits for all the \chandra data except for those from 2011 October and
2012 June are shown in Fig. \ref{fig:sn-dec10}. The best fit models for the
2011 October and 2012 June data are plotted in Fig~\ref{fig:sn-chandra}.

\subsubsection{XMM-Newton data}
\label{sec:xmm}

For the \xmm observations,  we use the best fit models  of UGC 5189A and the contaminating sources
 from the 2014 June  \chandra data  (nearest in time to \xmm observations) 
to remove the contamination in SN flux estimation. We again use an               
absorbed thermal plasma model to fit the SN spectra as described in
\S \ref{sec:chandra}.
 The data are best fit with 
$N_H(\rm CSM)=(2.64^{+0.69}_{-0.59})\times10^{21}$ cm$^{-2}$ with a reduced $\chi^2=1.41$.
We plot the spectrum and contour plots
in the lower panel of Fig. \ref{fig:nustar}.

\subsubsection{Swift-XRT data}
\label{sec:xrt}

The \swift  observations are usually closely spaced in time but
with  short ($<10$ ks) exposure times. To increase the
detection significance,  we combine several data
sets to extract the spectra near-simultaneously. 
Our criterion to group the spectra was to
get a uniform coverage on a logarithmic  scale. 
We, therefore, group the observations
taken during 2010~November~5-20, 2010~November~23-December, 
2011~April, 2012~October, 2013~January-March, 2013~May-June, 
2013~December, 2014~May-June,
and 2014~November-December.
The spectra were extracted using the online \swift spectrum extraction tool              
\citep{evan09,goad07}. 
We fitted the apec model with a fixed  temperature of 19 keV and 
 C-statistics were used.
In 2012 October  data (10.7 ks), 2013 December data (14.6 ks), and 2014 May--June  data
(13.1 ks), there are only 67, 46 and 35 counts, respectively, so we
fixed  $N_H(\rm CSM)$ to that of the best fit values 
closest in time, obtained from analysis of data of other telescopes.
 The \swift spectra are plotted in Fig. \ref{fig:swift}.

In the 2010 November  \swift data (Fig. \ref{fig:swift}), there is a clear indication of an extra feature around 6 keV, which was also seen in the
\chandra data around the same epoch.   
A Gaussian is best fit with an energy of $E_{\rm Gauss}=6.39^{+0.19}_{-0.23}$ keV. 
The unabsorbed 0.2-10~keV fluxes in the 
Gaussian and the continuum components are
$(3.39\pm1.28)\times10^{-13}$ erg cm$^{-2}$ s$^{-1}$
and $(26.15\pm6.67)\times10^{-13}$ erg cm$^{-2}$ s$^{-1}$,
respectively. In 2010 Dec data  we attempted to add a Gaussian, since it was also seen in the 2010 December \chandra data. However, the data have very few counts thus the         
fits
statistics do not change significantly.      

The best fit \sn models from our analysis
are listed in Table \ref{tab:analysis}. The 0.2--10~keV flux 
values are given in Table \ref{tab:sn-flux}.
In Fig. \ref{fig:sn3}, we plot the evolution of the 0.2--10~keV luminosity
as well as the column density. For the first 300 days, 
both quantities evolve slowly, but after 300 days, one can see a steep power law decline
(decay index $\sim 2$) in the luminosity.

 \begin{figure}
\begin{center}
\includegraphics[angle=0,width=0.40\textwidth]{f7a.eps}
\includegraphics[angle=0,width=0.40\textwidth]{f7b.eps}
\caption{In the top panel, we plot  the 0.2--10~keV light curve for SN 2010jl (Blue filled circles). We plot the bolometric light curve taken from \citet{fec+14}     
in blue squares. We also plot the X-ray light curve \citep{chandra12b} and bolometric luminosity \citep{strit12} of another well studied Type IIn supernova, SN 2006jd (orange circles and squares, respectively). Unlike SN 2010jl which has a very steep X-ray decay with index of $-2.12$, the SN 2006jd decline is  very flat (index $-0.26$) in a similar time range.   In the lower panel we plot the evolution of the CSM column density for SN 2010jl                          
as explained in Section~\ref{sec:sn}. 
}
\label{fig:sn3}
\end{center}
\end{figure}

 \begin{figure*}
\begin{center}
\includegraphics[angle=0,width=0.55\textwidth]{f8a.eps}
\includegraphics[angle=0,width=0.95\textwidth]{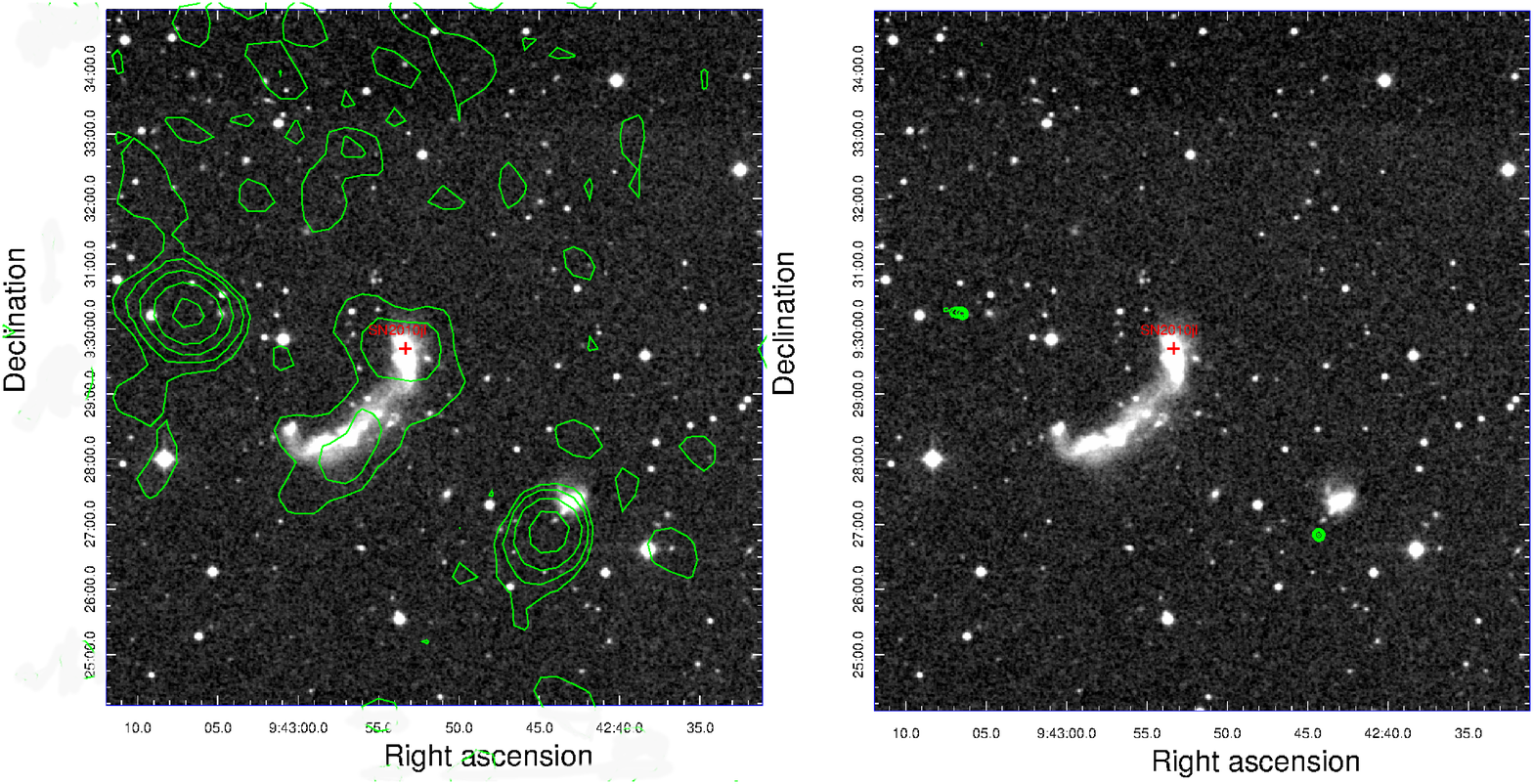}
\caption{\scriptsize{{\bf Upper Panel:}
Pre-explosion SN 2010jl FoV image with the VLA in the 1400 MHz band.  The flux units in the upper color bar are mJy.
 \sn is at a position $\alpha=09^h42^m53^s.337$, $\delta=+09^o29'42.''13$ (J2000). The immediate \sn region and UGC 5189A are
indistinguishable as the image resolution is $16''\times14''$.  
{\bf Lower panel:} The overlay NVSS ($45''$ 
resolution; left side) and FIRST ($5''$ resolution; right side) images on the J-band POSS II (Palomer 
Transient Survey II) image with the same contour levels.  Much of the 
extended emission is absent in the higher resolution image. 
}}
\label{fig:radio}
\end{center}
\end{figure*}

\section{RADIO ANALYSIS}
\label{sec:radio}

Before analyzing the \sn data, we examine the radio emission in
the SN FoV in Very Large Array (VLA)  
 archival data taken during 2006 December 1--21
in 1.4 GHz band. The telescope at the time of these observations was in
C-configuration.  The duration for all the datasets ranged from
30 minutes to 3 hours (including the calibrator). We carried out a joint 
analysis of the data.                   
The image resolution obtained was $16''\times14''$ and the map rms was
 191 $\mu$Jy (Fig. \ref{fig:radio}). The figure 
shows that  SN 2010jl lies in a region of extended radio emission.
To investigate  the nature of the radio emission, we  looked into the
VLA NVSS (resolution $45''$) and FIRST (resolution $5''$) images of the SN FoV  in the 1.4 GHz band.  In Fig. \ref{fig:radio}, we
overlay the Second Palomar Observatory Sky Survey (POSS II) J-band gray image with the
NVSS and the FIRST contours.
The figure suggests that the extended emission is resolved at higher resolutions and not likely to contaminate the SN flux.

The first radio data of SN 2010jl  were taken on 2010 November 6.59 UT in
EVLA C-configuration at 8 GHz band. The total duration of the observation 
including overheads was 30 minutes. The data quality was good and only 
7.5\%  of the data were flagged. The map rms in the 8 GHz band              
was 24 $\mu$Jy and the synthesized 
beam size was $2.35'' \times   2.07''$. We did not detect  \snb. The flux density
at the \sn position was $-41\pm24$ $\mu$Jy.
We then attempted to observe the  \sn in 33 GHz band to account for
the scenario in which the SN was
absorbed at lower frequencies.  The observations were taken on 2010 November
8.47 UT for 3588.7 s. We obtained a rms of 45 $\mu$Jy and image resolution of
$0.67'' \times   0.60''$. We did not detect  \sn in this band either. The flux
density at 33.56 GHz at the \sn position was $52\pm45$ $\mu$Jy.

We continued to observe \sn at regular intervals. 
The first detection of the SN came on 2012 April 18            
in the 22 GHz band, with a flux density of $60.9\pm17.6$ $\mu$Jy.  
To estimate the flux density of  \sn in all the images, we fit two  Gaussian models, one for
the SN component and one for the  underlying background level to take
care of any underlying extended emission. 
Since then we have
been detecting  \sn in various VLA bands (Table \ref{tab:radio}). 
Our most secure detections are
in 2012 December, when the VLA was in the A-configuration.

\begin{figure}
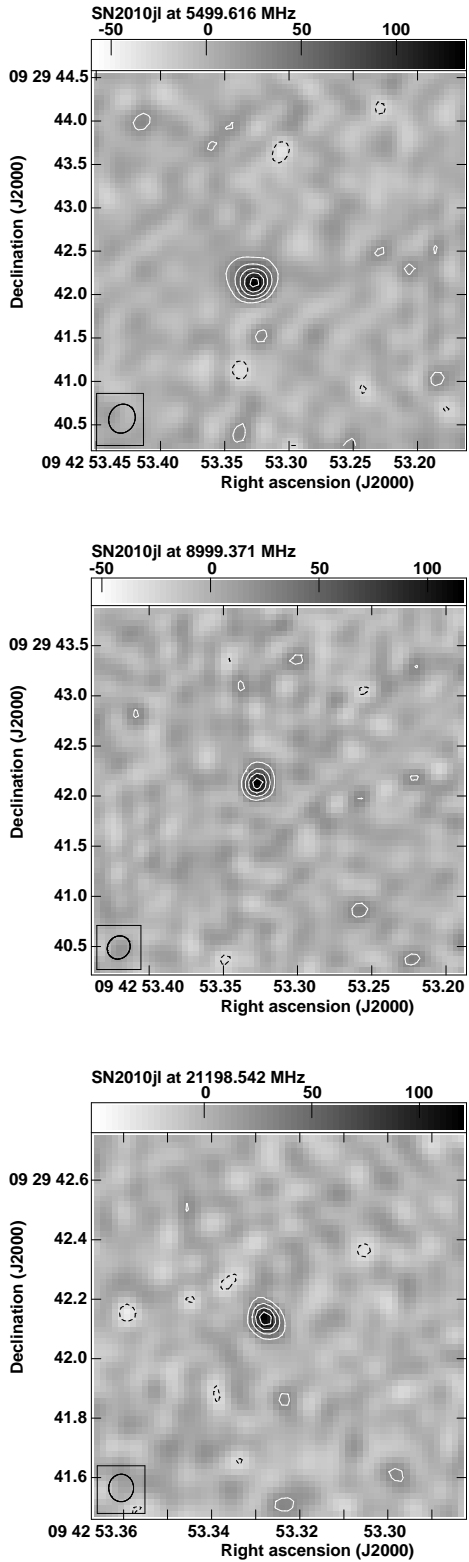

\centering
\includegraphics[width=0.35\textwidth]{f9a.eps}
\includegraphics[width=0.35\textwidth]{f9b.eps}
\includegraphics[width=0.35\textwidth]{f9c.eps}
\caption{VLA A-configuration 5 GHz, 8 GHz and 21 GHz detections of SN
2010jl, taken on 2012 Dec 01.46, 02.40 and 02.53, respectively. The units of the color bars above the maps are  $\mu Jy$.}
\label{radio}
\end{figure}

 \begin{deluxetable*}{cccccccc}
\tabletypesize{\scriptsize}
\tablecaption{VLA radio observations of SN 2010jl
\label{tab:radio}}
\tablewidth{0pt}
\tablehead{
\colhead{Date of} & \colhead{Days since}   &
\colhead{Config.} & \colhead{Central} & 
\colhead{BW\tablenotemark{b}} & 
\colhead{Resolution}
 & \colhead{Flux density\tablenotemark{c}} & \colhead{rms} \\
\colhead{obsn. (UT)} &
\colhead{Explosion\tablenotemark{a}} &
\colhead{} & \colhead{GHz} & \colhead{Freq, GHz} & \colhead{($''\times''$)}
 & \colhead{$\mu$Jy} & \colhead{$\mu$Jy}
}
\startdata
2010 Nov 06.60  & 37.60 & C & 8.46 & 0.26  & $2.35\times2.07$ & $<71.1$ &23.7\\
2010 Nov 08.48  & 39.48 & C & 33.56    & 0.26  & $0.67\times0.60$ & $<133.5$ & 44.5\\
2010 Nov 08.52  & 39.52 & C & 22.46 & 0.26 & $0.93\times0.85$ & $<74.4$ & 24.8\\
2010 Nov 14.52 & 45.52 & C & 22.46 & 0.26 & $0.93\times0.86$ & $<82.8$ & 27.6\\
2010 Nov 23.43  & 54.43 & C & 22.46 & 0.26 & $1.06\times0.92$ & $<70.1$ & 23.4\\
2011 Jan 21.26  & 113.26 & CnB & 7.92 & 0.13 & $3.71\times0.84$ & $<76.8$& 25.6\\
2011 Jan 21.26  & 113.26 & CnB & 4.50 & 0.13 & $6.52\times1.51$ & $<116.7$ & 38.9\\
2011 Jan 21.39  & 113.39 & CnB & 22.40 & 0.26 & $0.98\times0.78$ & $<107.4$&35.8\\
2011 Jan 22.26 & 114.26 & CnB & 22.46 & 0.26 & $1.03\times0.37$ & $<79.8$&26.6\\
2011 Jan 23.27  & 115.27 & CnB & 4.50 & 0.13 & $5.26\times1.63$ & $<76.2$ & 25.4\\
2011 Jan 23.27  & 115.27 & CnB & 7.92 & 0.13 & $2.93\times0.91$ & $<55.5$ & 18.5\\
2011 Apr 22.25  & 204.25 & B & 22.46 & 0.26 & $0.41\times0.28$ & $<114.0$ & 38.0\\
2011 Jul 07.99  & 280.99 & A & 4.50 & 0.13 & $0.60\times0.42$ &$<78.0$ & 26\\ 
2011 Jul 07.99  & 280.99 & A & 7.92 & 0.13 & $0.27\times0.22$ & $<62.1$ & 20.7\\
2012 Jan 24.28 & 481.28 & DnC$\rightarrow$C & 8.46 & 0.26 & $3.35\times2.46$ & $<123$&23.4\\
2012 Feb 28.20 & 516.20&C& 8.55 & 1.15 & $2.65\times2.16$ & $<32.1$ & 10.7\\
2012 Mar 02.22  & 519.22 & C & 5.24 & 1.54 & $3.89\times3.53$ & $<33.6$ & 11.2\\
2012 Apr 15.03  & 563.03 & C & 8.68 & 1.41 & $2.77\times2.25$ & $<45.6$ & 15.2\\
2012 Apr 15.05  & 563.05 & C & 5.50 & 2.05 & $3.89\times3.29$ & $<32.7$ &10.9\\
2012 Apr 18.03 & 566.03 & C & 21.20 & 2.05 & $1.13\times0.98$ & $60.9\pm17.6$ & 10.1\\
2012 Apr 22.10 & 570.10 & C & 21.20 & 2.05 & $1.08\times0.87$ & $38.3\pm20.7$ & 13.5 \\
2012 Aug 11.71  & 681.71 & B & 9.00 & 2.05 & $0.84\times0.71$ & $76.6\pm20.3$ & 12.9\\
2012 Aug 12.71  & 682.71 & B & 5.50 & 2.05 & $1.26\times1.08$ & $111.9\pm17.8$ &12.6\\
2012 Dec 01.46  & 793.46 & A & 5.50 & 2.05 & $0.34\times0.30$ & $131.3\pm 22.1$ & 11.1\\
2012 Dec 02.40  & 794.40 & A & 9.00 & 2.05 & $0.24\times0.22$ & $118.8 \pm 16.8$ & 9.8\\
2012 Dec 02.53  & 794.53 & A & 21.20& 2.05 & $0.09\times0.08$ & $115.3 \pm 17.1$ & 9.9\\
2013 Jan 18.34 & 841.34 & A$\rightarrow$D & 21.20 & 2.05 & $0.17\times0.05$ & $<109$& 10.7\\
2013 Jan 18.38  & 841.38 & A$\rightarrow$D & 9.00 & 2.05 & $0.32\times0.12$ & $<470$ & 20.9\\
2013 Jun  10.96  & 984.96 & C & 8.68 & 2.05 & $2.31\times2.03$ & $123.0\pm26.8$ & 9.7\\
2013 Jun 11.94  & 985.94 & C & 5.50 & 2.05 & $3.98\times3.23$ & $91.3\pm34.8$  & 12.7\\
2013 Aug 10.74  & 1045.74 & C & 21.20&2.05 & $1.09\times0.90$ &$<69.3$  & 23.1\\
\enddata
\tablenotetext{a}{Assuming 2010 October 1 as the explosion date \citep{stoll11}}
\tablenotetext{b}{Bandwidth of the observation}
\tablenotetext{c}{Since SN is off the Galactic
plane, the errors in the SN flux due to calibration errors will be less than 5\%. In  case of non-detections, the 3-$\sigma$ flux density limit is quoted.}
\end{deluxetable*}

To determine the \sn position, we have used 2012 December 2 data in 22
GHz  band 
when the  VLA was in A-configuration.  In this data, we obtained a  resolution of $0.09''\times0.08''$. 
 The best SN position is
$\alpha=09^h42^m53.^s32773\pm0.00021$, $\delta=+09^o29'42.''13330\pm0.00344$ 
(J2000), which agrees well within 0.15$''$ accuracy with
the optical  position given by \citet{ofek14}.

The post detection observations of \sn with the VLA  C-configuration in 2013 June              
were contaminated by the
extended flux due to poorer resolution in this configuration, especially in the 5 and 8 GHz bands. 
In this case, we have used the C-configuration
images of 2012 February 
before the \sn detection in the respective frequencies and subtracted it from the
post detection images to get the uncontaminated flux of  \snb.

In Fig. \ref{radio}, we plot the  contour plots of SN 2010jl in the 5 GHz, 8 GHz and 21 GHz bands for this epoch. 
The SN flux densities  obtained in 2013 Jan when the VLA was in A$\rightarrow$D configuration are not very reliable 
 due to contamination from the underlying extended emission.

\begin{figure*}
\centering
\includegraphics[width=0.95\textwidth]{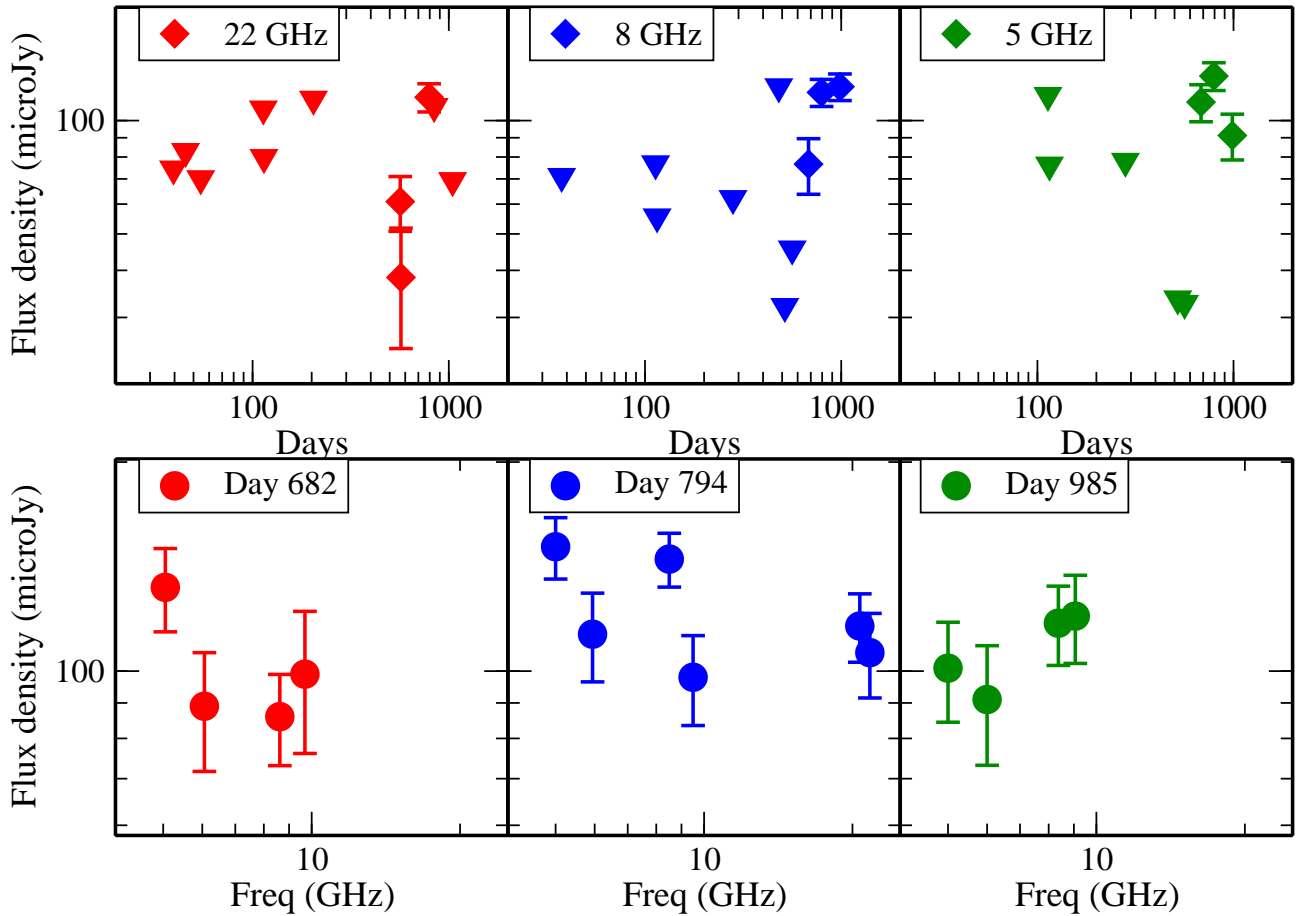}
\caption{Radio light curves of 
\sn are shown in the upper panel. The  inverted triangles are  $3\sigma$ upper 
limits. 
The lower panel shows the radio spectrum at 3 epochs.
}
\label{lc}
\end{figure*}

In Fig. \ref{lc}, we plot the light curves of 
the SN (upper panel), and in the lower panel  plot the spectra at 3 epochs when the SN was detected in multiple bands. For the spectra
with the JVLA data, we have divided the 2 GHz bandwidth into 2 subbands and imaged it independently to get the flux densities in the
two subbands. 

\section{DISCUSSION}
\label{sec:discussion}

Using the \nustar data, we derive a shock temperature of 19 keV, which is 
consistent with the analysis of \citet{ofek14}. This corresponds to 
a shock velocity of $\sim 4000$ $\kms$ at around 750 days. At earlier epochs, the shocked gas may be hotter, but we cannot determine the exact temperature at
any other epoch due to the absence of hard X-ray observations.
This may introduce some errors in column density estimate and the SN flux. We attempt to quantify this error. In a standard SN-CSM interaction model, the temperature of the shock varies with time $t$
as $t^{-2/(n-2)}$ (where $n$ is the power-law index of the ejecta density profile), i.e. $t^{-0.25}$ for a typical value $n=10$ \citep{cf03}. Thus 
over the full span of our observations (day $\sim$50 to 
$\sim$1500), the temperature 
would change by a factor of $\sim2$. In our fits, we estimate the 
change in column density and in SN flux due to the change in temperature. We find that for a 20\% change in temperature, the column density changes by 5~\% and the SN flux changes by $1.5$~\%. If we change the temperature of the shock by 100~\%, the change in column density is
less than 15~\% and in the SN flux is less than 10~\%. This analysis
shows that even though we do not have a handle on the temperature at various epochs and are assuming a constant temperature,
the errors introduced in other parameters like column density and
the SN flux do not suffer significantly.

\subsection{Summary of main results}
\label{sec:results}

SN 2010jl 
is the only Type IIn  SN for which a well sampled X-ray dataset exists.
Thus we are able to trace the evolution of column density and light curve
 all the way from 40 to 1500 days (Fig. \ref{fig:sn3}).

  The X-ray luminosity for \sn is roughly constant for the first $\sim 200$ 
days with a power law index of $0.13\pm 0.08$. Unfortunately there are no data 
between 200 to 400 days, but from day 400  onwards
 one can see a faster decay in the luminosity  following a powerlaw index of $-2.12\pm0.13$. 
 For a shock velocity  of 
 4000 km s$^{-1}$ estimated above, 
 the start of the rapid decline in X-rays corresponds to a radius of
 $1.3\times10^{16}$ cm.
 In  Fig. \ref{fig:sn3}, we overplot the bolometric luminosity lightcurve of \sn
  taken from \citet{fec+14} on the X-ray light curve of the SN.
We note that the bolometric luminosity also declines 
rapidly  around day 300, consistent with the X-ray lightcurve.
\citet{fec+14} have argued that the steepening in the 
bolometric luminosity coincides with the H$\alpha$ shift becoming constant and  have explained it a result of less efficient 
  radiative acceleration. 
 \citet{ofek14} argue that this is the time when the SN reached the 
 momentum conserving snow-plow phase.   However, as we will show  in
 Section~\ref{sec:model}, 
this can be explained in our simple model by  changes in the CSM density profile.
 
 In Fig. \ref{fig:sn3}, we also plot the X-ray and bolometric luminosities of 
another well sampled Type IIn, SN 2006jd \citep{chandra12b,strit12}. The  luminosity decline is
 much flatter in SN~2006jd than that of SN 2010jl.   Around the same
 epoch, the  SN 2006jd X-ray lightcurve 
 declines as $t^{-0.24}$, while SN 2010jl declines as
 $t^{-2.12}$. 
The relative flatness of the bolometric luminosity of SN 2006jd for a longer duration indicates that the CSM interaction powered the light curve for a 
 much longer time than  \sn. 
This indicates that  the duration of mass ejection in 
 SN 2006jd may have been longer in this case than for SN 2010jl, though  in both cases 
 occurred   shortly before the explosion. This suggests different nature of progenitors for the two SNe.

The most interesting result of the paper is 
the evolution of the column density with time. 
At $t \sim 40$ days, the column density is 3000 times higher than the Galactic column 
density, and declines by  a factor of $\sim 100$ by the epoch of our last observation in 2014 Dec.  
Since the higher column density is not associated with the high host
galaxy extinction, this indicates that the higher column density is due to the 
CSM in front of 
the the shock where the dust is evaporated,
thus is arising
from the CSM. 

The presence of broad emission lines seen in early \sn optical spectra \citep{smith12, fec+14, ofek14} can be explained by electron scattering \citep{chugai01}.                 
 This requires an electron scattering optical depth  $> 1-3$, i.e. a column density       
greater than  $3\times10^{24}$ cm$^{-2}$.                                                 
 Comparable values of the column density were seen only in the first X-ray observations   
($t < 70 $ d; Table \ref{tab:analysis}). At later epochs ($t > 70 $ d) we only                                      
  see the X-rays which pass through a column  density $<10^{24}$ cm$^{-2}$. These         
constraints are difficult to  reconcile with a spherical model of the column density      
$\sim 3\times10^{24}$ cm$^{-2}$.                                                          
We therefore should admit that either the width of emission lines at the late time is     
not related to  Thomson scattering, or the X-rays escape the interaction region        
avoiding the CSM with a high column density.
Thus, a possible scenario for the CSM in \sn is the same          
as the bipolar geometry in the CSM of 
$\eta$--Carinae \citep{smith06}.

An intriguing issue here is the presence of an extra component
 in the 2011 October  and 2012 June  \chandra data, which is not present 
 before or after. This component is well fit with a power law  index of 1.7, though with a varying column density, with much higher column density at the
2011 October  epoch. The origin of this component is not clear. However, the component seems to spatially coincide with the the position of the SN
and occurs when there are changes to the SN spectrum in the energy range close to that of the extra component.  This would 
seem to suggest that the emission is related to the SN.  
One possibility is that the soft component is a result of a cooling shock. In this scenario a
 mass loss rate of $0.1 M_\odot$yr$^{-1}$ \citep{fec+14}, 
wind velocity $\sim 100$ $\kms$ \citep{fec+14}, and shock temperature of 19 keV (this paper)
corresponds to a cooling time of 86 days at $10^{16}$ cm 
(roughly the radius at one yr for ejecta velocity of 4000 $\kms$).  
Thus the forward shock 
should be cooling around this time, as mentioned in our discussion of 
the optical light curve. An adiabatic shock model underestimates the low 
energy X-ray flux. 
The fact that the need for an additional component disappears could 
be a result of the shock becoming adiabatic. 

While the radio observations of \sn started as early as 
$\sim40$ days post
explosion, the first radio detection was around day 566.   
Unlike its X-ray 
counterpart,
the radio luminosity from \sn is weak for a SN IIn  (Fig. \ref{lc}).
This is unlike  SN 2006jd which was an order of magnitude 
brighter in radio bands than SN 2010jl \citep{chandra12b}.
Like most of the known radio Type IIn SNe, SN 2010jl rises at
radio wavelengths at late times, most likely due to  absorption 
by a high density of
 CSM, or
due to internal absorption \citep[e.g.,][]{chandra12b}.
We have not attempted to fit a detailed model to the radio data in view 
of the small number of detections and small range of time.
However, the existing data have distinctive features as seen in 
Fig.\  \ref{lc}. The light curves are
fairly flat, as are the frequency spectra.
For the standard models of radio SNe, these properties occur when 
the SN is
making the transition from optically thick to optically thin 
\citep[e.g.,][]{weiler02,cf03}.  Thus, most likely we have detected the radio emission near the peak of the synchrotron emission.
  
     From Fig. \ref{lc}, we estimate that  the light curve peak                  
at 5 GHz occurs on about day 900 at a flux of $F_{ob}\approx 0.12$ mJy,
or a luminosity of $3.4\times 10^{27}$ ergs s$^{-1}$ Hz$^{-1}$.
The time of maximum is typical of SNe IIn, but the 
luminosity is lower than most,
suggesting that the absorption mechanism is not 
synchrotron self absorption if
the radio emitting region is expanding at $4000 - 6000\kms$ 
as indicated by
X-ray \citep[this paper and ][]{chandra12a} and near IR observations
\citep{borish14}. This can be seen in  Fig.\ 1 of \citet{chevalier09}.
In the case of the Type IIn SN 1986J, the radio expansion was measured 
by VLBI techniques and an
expansion velocity of $5700\pm 1000\kms$ was found over 
the period 1999-2008
\citep{bieten10}.
Thus an expansion velocity of $5000-6000\kms$ is plausible. 
In this case, the
absorption would have to be something other than
 synchrotron self-absorption,
most likely free-free absorption.  
The fact that the 8 GHz radio flux appears around day 700 implies
the emission measure along the line of sight at this stage
$\langle n_e^2\ell \rangle \sim 8\times10^{26}$ cm$^{-5}$, 
where $\ell$ is the linear size of the absorbing gas region and     
assuming $T_e=10^4$ K.
On day 700 the radius of the shell with an average velocity of 
4000 km s$^{-1}$ is of $r\sim2.4\times10^{16}$ cm. Adopting
$\ell \sim r$ we thus come to the rough estimate of the column density
$N_{H} \approx 4\times 10^{21}x^{-1}$ cm$^{-2}$, where $x$ is the 
hydrogen ionization fraction. 
A low hydrogen ionization fraction of $x \sim 0.1$ is needed 
for the value of $N_H$ to be consistent with
the column density on day 700 recovered from the X-ray data
($3\times10^{22}$ cm$^{-2}$).
Although we expect the CSM to be ionized at early times, 
recombination is possible at later times; however,
a detailed study of the ionization balance in the CSM of 
SN 2010jl is beyond the scope of this paper.

Another possibility for explaining the escape of the radio emission 
is that it  comes from a different region than the X-ray emitting
region.  To have a lower column,
the radio region would probably come from expansion into a lower 
density part of the CSM
and would thus be more extended.  The X-ray emission from
the radio region may not be detected because of the low density.  
Again, details are beyond the
scope of this paper.

\subsection{Circumstellar interaction modeling}
\label{sec:model}

The detailed evolution of the absorbing column density derived
from the X-ray data provides us with an unprecedented opportunity
to examine the CSM around a SN~IIn in both X-ray and optical bands.
The question  arises of whether the optical light
curve powered by the CSM interaction is consistent with the observed
column density. Here, we present a simple
circumstellar interaction model which suggests that freely expanding SN
ejecta collide with the dense CSM. In a smooth dense CSM the interaction 
zone consists of a forward and a reverse shock along with a cool dense 
shell (CDS) formed in-between. 
 We confine ourselves to the interaction 
hydrodynamics based on the thin shell approximation \citep{chevalier82}.  
The equations of motion and mass conservation are 
integrated for arbitrary density distributions in both ejecta and 
CSM using a 4th order Runge-Kutta scheme. The model provides us with 
the CDS radius ($R_s$) and velocity ($v_s$),
the forward shock speed $v_{fs} \approx v_s$, the reverse shock speed 
$v_{rs} = (R/t - v_s)$, and the kinetic luminosity released in the 
shock $L_j= 2\pi R_s^2\rho_j v_j^3$, where $\rho_j$ is preshock density 
and  $v_j$ is the shock velocity ($j=rs, fs$).

Generally, the conversion of the kinetic energy into the 
radiation in the interaction SNe is affected by  
complicated hydrodynamic and thermal processes including the thin shell 
instability \citep{vishniac94}, the Rayleigh-Taylor instability of the 
decelerating CDS, the CSM clumpiness,
mixing, and energy exchange between cold and hot components
via  radiation and  thermal conductivity.                      
This makes the computation of the radiation output of the CS interaction 
quite a formidable task. We use a simple approach 
in which the X-ray luminosity of both shocks is 
 equal to the total kinetic luminosity times the radiation efficiency
 $\eta = t/(t+t_c)$, where $t_c$ is the
 cooling time of the shocked CSM at the age $t$.
To calculate the shock cooling time we assume a constant postshock 
density four times the upstream density $\rho_0$, 
while the shock temperature 
is calculated in the strong shock limit.
We find that the reverse shock is always fully radiative in our models.
 In the case of SN 2010jl, the optical luminosity dominates the 
observed X-ray luminosity by a factor of ten. 
In our model, the bolometric luminosity, which is primarily in the optical, is equal to the total radiation luminosity.

\begin{figure}
\centering
\includegraphics[width=0.45\textwidth]{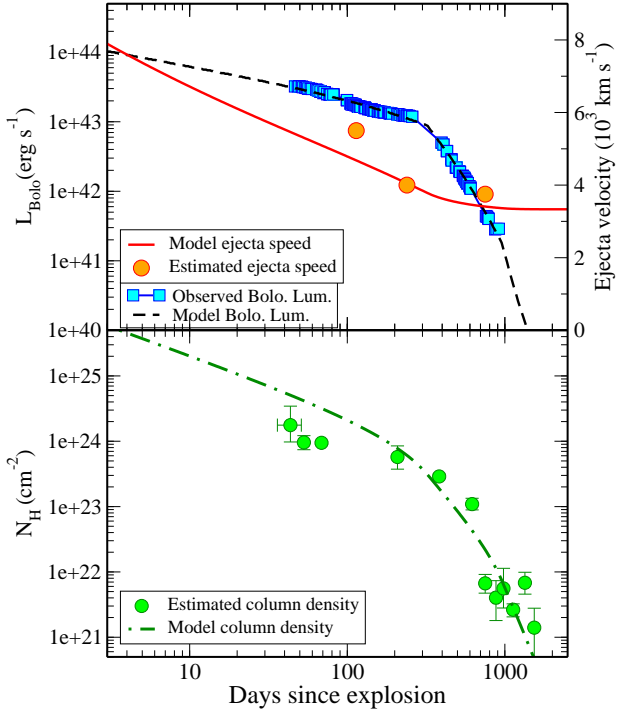}
\caption{\scriptsize{
 {\it Top Panel}: The 
bolometric light curve obtained from our modeling detailed in \S \ref{sec:model}
({\it dashed} line) plotted over the observational data
({\it blue squares}) which are taken from \citet{fec+14}.  The  {\it  red solid} line is the  shell velocity obtained from our modeling
(in units of $10^3$ km s$^{-1}$) compared to the observational data 
({\it filled circles}) taken at various epochs \citep[this paper on day 750 and][at earlier epochs]{borish14}. {\it Lower Panel}: the CSM column density outside 
the forward shock obtained from our modeling in \S \ref{sec:model} versus the data recovered from X-ray observations
({\it squares}). Our simple model detailed in \S \ref{sec:model} is able to reproduce the
observational quantities quite well.
}}
\label{fig:opt}
\end{figure}

The model assumes that the optical radiation generated in the
interaction zone instantly escapes the CSM which means 
that the diffusion
time \citep[$t_{dif} \sim (r/c)\tau$, where $c$ is the speed             
of light and and $r$ is the shell radius, ] 
[]{chev81} is smaller than the expansion time, or the CSM optical depth
$\tau < c/v$ (where  $v$ is the shell speed). In our model, on day 3
the column density of the wind is $5\times10^{25}$ cm$^{-2}$  and the
Thomson optical depth at this stage is  therefore 30. This value is 
comparable to $c/v \sim 30$ assuming $v=10^4$ $\kms$. 
Therefore, after about day 3 the diffusive trapping of photons in 
the wind can be ignored. The SN ejecta density distribution
is approximated by the analytical expression 
$\rho_e\propto 1/[1+(v/v_0)^n]$
which reflects an inner plateau ($v<v_0$) and an outer power law density
drop with $7\leq n <10$. 
The  initial ejecta boundary velocity $v_b$ is assumed to be  
$3\times10^4$ km s$^{-1}$; our results are not sensitive to 
the boundary velocity in the range of $(1-3)\times10^4$ km s$^{-1}$ 
at epochs $t > 5$ d.  
The CSM density distribution is set by a broken power law
$\rho \propto r^{-m}$ with $m=m_1=2$ in the range of $r<r_1$ and
$m=m_2>2$ for $r>r_1$. 
The value $m_1=2$ was chosen for small radii because it is 
an approximate fit and is plausible for a wind; the outer
value of $m_2$ is a 
fitting parameter.
The ejecta diagnostics based on the CSM interaction cannot
constrain the ejecta mass and energy uniquely because the same
density versus velocity distribution in the ejecta outer layers can be
produced by a different combination of mass and energy. Yet the observed
interaction luminosity and the final velocity of the decelerated shell
($v_f$) constrain the energy of the outer ejecta with the velocity $v>v_f$.      
For the adopted density power law index $n$ the obvious relations              
$v_0 \propto (E/M)^{1/2}$, $\rho(v) \propto \rho_0(v_0/v)^n$, 
and $\rho_0 \propto M/v^3$ result in the energy-mass
scaling $E\propto M^{(n-5)/(n-3)}$, or $E\propto M^{0.6}$,
assuming $n=8$. In turn, 
this scaling, when combined with the requirement
that the velocity $v_f$ should be larger than the ejecta 
density turnover
velocity $v_0$, provides us with the lowest plausible values of $E$ and $M$.
As a standard model we adopt an ejecta mass of $8~M_{\odot}$    
which barely satisfies the condition $v_f > v_0$.
The corresponding kinetic energy of ejecta is then
$2.1\times10^{51}$ erg.

The optimal model is found by the $\chi^2$-minimization            
in the parameter space of $r_1$, $m_2$, and the wind density parameter 
$w_1=4\pi r^2\rho$ in the region $r<r_1$.
We found the best-fitting values 
$w_1=4.1\times10^{17}$ g cm$^{-1}$, $r_1=1.29\times10^{16}$ cm, 
and $m_2=4.45$. The parameter errors determined via
 $\chi^2$ variation do not exceed 1.5 \%. These errors 
are formal and too optimistic, since we ignore uncertainties 
in the distance 
and the extinction, and the systematic error related to the 
model assumptions. The model (Fig. \ref{fig:opt}) 
reproduces the \sn bolometric light curve \citep[obtained 
from][]{fec+14} 
quite well and produces reasonble values for the shell velocity,
consistent with the observational estimates \citep[obtained from this 
paper on day 750 and][for the earlier two epochs]{borish14}.
The evolution of the column density obtained from X-ray data
is also described by our model at $t\geq 200$ d as well, except 
for the early epoch 
$t < 100$ d when  the model requires a somewhat larger           
column density.
In the model, the CSM density power law index breaks at
$r_1=1.3\times10^{16}$ cm from $m=2$ to $m=4.45$, which suggests
that the bulk, $2.6~M_{\odot}$, of the total CSM mass, $3.9~M_{\odot}$,
lies within the radius $r_1$. In order to describe the column density
at  late epoch $t>600$ d, the power law index should become steeper
($m\approx 7.5$) for $r>r_2=3.2\times10^{16}$ cm.  A lower density region 
outside the close-in CSM is consistent
with what is deduced by \citet{fec+14} in their analysis
for the narrow lines.

The agreement between the optical model CSM column density and that
inferred from X-ray data suggests that the CSM density 
recovered in the model is realistic. The external radius of the CSM envelope,  
$1.3\times10^{16}$ cm, combined with  
the wind velocity of 100 km s$^{-1}$ implies that the SN event
was  preceded by an episode of  vigorous mass loss 
starting 40 yr prior to the explosion. The mass loss rate 
at this stage was $\sim0.06~M_{\odot}$ yr$^{-1}$, which
 is  similar to the value obtained in \citet{fec+14}.

Although the overall picture of the X-ray generation by CSM interaction 
with  absorption in the CSM is generally convincing, 
there is an issue with the X-ray luminosity.
The point is that the unabsorbed X-ray luminosity
is significantly (a factor of 10) lower then the optical.  
But at late times, $t>100$ d, the forward shock
dominates  the luminosity so
one would expect that more than half of the total X-ray radiation
 escapes the interaction zone, at odds with the observations.
The disparity suggests there is some mechanism
that  converts the kinetic luminosity into the optical radiation
avoiding significant hard X-ray ($h\nu > 1$ keV) emission. 
 The soft XUV radiation then could be absorbed by the CSM resulting in
the X-ray deficit. 
Given the Thomson optical depth $\tau_T \lesssim 1$, 
$h\nu/mc^2 \ll 1$, and $4kT_e/mc^2 \ll 1$, neither  Compton scattering
in the CSM nor  Compton cooling of hot electrons in the forward
shock are able to provide the required degree of X-ray 
softening \citep{chevalier12}.  
An alternative scenario is conceivable which connects the X-ray deficit
to CSM clumpiness. 
Although radiative properties of the shocked CSM
cannot be reliably quantified, the predominance of  soft radiation
is a likely outcome in this case.
Indeed, if the bulk of the CSM were
in clumps the luminosity of the shocked intercloud gas would be weak.
The cool matter mixed with hot gas in this scenario
becomes a dominant source of the radiation that falls into the soft XUV
band.  Thermal conductivity might allow heat flow from the
hot shocked intercloud gas into the mixed cool gas.
Alternatively, \citet{fec+14} have  explained this deficit to be due to the
  presence of an anisotropic CSM, because of which most of the X-rays 
  are absorbed and converted into optical photons.
We note that  in the case of SN 2006jd, the
bolometric luminosity is
around an order of magnitude larger than the X-ray luminosity
(Fig. \ref{fig:sn3}).

\subsection{Comparison with Other Results}

In our X-ray spectral fits of UGC 5189A and other nearby sources, we allow 
the column density to vary freely and also use a metallicity of 0.3 
\citep{stoll11} for the excess column density (over Galactic). 
We derive a host column density of $\sim 4\times 10^{21}$ cm$^{-2}$, whereas, \citet{ofek14} have fit all the nearby X-ray sources 
 with a
fixed Galactic column density ($3\times10^{20}$ cm$^{-2}$) and a power law 
spectrum with photon index $\Gamma=1.375$.  Part of the discrepancy can be accounted for the fact they have used solar metallicity as opposed to 0.3 solar metallicity for the host Galaxy used by us. This will lower their equivalent Hydrogen column density in the XSPEC fits by
a factor of three.
\citet{fec+14}, using a fit to the Lyman-$\alpha$ damping wings,
find $N_H(\rm Host)=(1.05\pm0.3)\times10^{20}$ cm$^{-2}$ for the
host galaxy, while the corresponding
value is $N_H(\rm Galactic)=(1.75\pm0.25)\times10^{20}$ cm$^{-2}$
from the Milky  Way. Their
Galactic column density is  around a factor of 2 lower than the one derived from \citet{dl90}. However, the GMRT has observed 
the \sn host galaxy UGC 5189 in 
21 cm radio bands  \citep{jayaram14} in 2013~November-December, and they
derive the HI column density  to be $2.4\times10^{21}$ 
cm$^{-2}$, consistent with our best fit values.
We  caution here that the column densities obtained in our fits have large uncertainties, and our estimates for the host galaxy may be treated as an upper limit on the host column density.

Although our best fit temperature agrees with that found by \citet{ofek14}, our column density is somewhat  smaller ($N_H(\rm        CSM)=(6.67^{+2.47}_{-1.94})\times10^{21}$ cm$^{-2}$) than that quoted by        
\citet{ofek14} which is $\sim10^{22}$ cm$^{-2}$; the difference is 
more significant considering they have used solar metallicity. 
To test the robustness of our best fit column density we have 
attempted to fit only the \xmm spectrum as \xmm data are especially sensitive to the column density due to absorption being dominant in the  0.2--10~keV energy range. We obtain a column density                              
of $N_H(\rm CSM)=(6.67^{+2.36}_{-1.86})\times10^{21}$ cm$^{-2}$
($\chi^2=1.07$), which is consistent with our joint fit.
The
contour plot of the column density for \xmm data shows that
it is well constrained  (Figure \ref{fig:nustar}).

\citet{ofek14} dispute  the values of high column density 
at early times obtained by \citet{chandra12a} by claiming that
they have used  many parameters. However,
in addition to the fits to X-ray data, the additional evidence of a high column density comes from 
the Fe~K-$\alpha$ line seen in 2010 December 7--8 \chandra data  and 2010 November \swift data, which suggests
$N_H(\rm CSM)=2 \times 10^{24}$ cm$^{-2}$ (assuming $Z/Z_\odot=0.3$ and 
$EW=0.2$ keV),
 consistent with the absorption column density obtained in our fits.

 \citet{ofek14}  fit the early \chandra data
with blackbody models, considering the medium to be optically thick to X-rays. However, 
a blackbody fit corresponds to an extremely small emitting area. For example, at an early 
epoch, for an X-ray  luminosity of $8.5\times 10^{41}$ erg s$^{-1}$
and a temperature of                           
3.4~keV, the blackbody emitting radius is only $~2\times 10^{7}$ cm, 
which is physically not plausible, considering the expected large area of the shock front. 
Thus we disfavor the blackbody model. 
\citet{ofek14} also fit their models using a powerlaw. In our models, we
have tried to fit the data using powerlaw models. While a powerlaw does give acceptable fits, the photon index is very flat ($\Gamma \le 0.5$), which is physically implausible.

The steepening in the X-ray and bolometric luminosity light curve around day 300 is quite significant. \citet{ofek14} explain the  steepening by arguing that the shock reached the fast cooling snow-plow phase. However, in our model we can
explain this  by introducing a steepening in the
density profile. 
Our mass loss estimates are a factor of 10 smaller than \citet{ofek14}, but
are consistent with \citet{fec+14}. This discrepancy can be partly accounted 
by the fact that \citet{ofek14} have assumed a wind velocity of 300 $\kms$,
whereas we assumed it to be 100 $\kms$, adopted from \citet{fec+14}. 
   
\section{CONCLUSIONS}

Here we have reported the most complete  X-ray and radio observations
  of a luminous Type IIn supernova SN 2010jl.     SN 2010jl is the only Type IIn SN which has been well sampled in
 both radio and X-ray bands since early on. 
 
 Using publically
available \nustar data, we determine a temperature for the shocked gas that  is consistent with the value obtained from \citet{ofek14}. 
 The 6.4 keV Fe-K$\alpha$ line seen in the first \chandra epoch \citep{chandra12a} is also present in the \swift data 
 taken around the same time, confirming that this line is real.
 
 While the radio emission is  weak in \sn, 
 the X-ray luminosity is one of the highest for  a Type IIn SN.         
 This provides us  a unique opportunity to trace the
 evolution of the circumstellar column density. 
 The circumstellar column densities at various epochs are 10--1000 
 times higher  than that of the host galaxy. 
 This evolution is satisfactorily reproduced in
a model in which freely expanding SN
ejecta collide with the dense and smooth CSM, with the                
 forward shock luminosity being a fraction of
 the kinetic luminosity depending upon the radiation efficiency. 
 
 The X-ray light curve evolution is quite flat for the first 200 days. 
 However, after $\sim 400$ days, the light curve shows a steep decline. 
There is a similar steepening of the optical luminosity, from which we infer
a steepening of the CSM density power law index from 
the standard $r^{-2}$ profile.
In contrast, the Type IIn  SN 2006jd light curve at a similar epoch is  much flatter,           
 indicating  differences in the mass loss leading up to the SN. 
 The case of SN 2010jl  is  consistent with rapid mass
 loss beginning a short time before explosion.

The observed radio emission for \sn is very weak and does not clearly evolve as in standard models.
The radio spectra and their evolution suggest that the emission is close to its peak
at an age of $\sim 10^3$ days.
The implication is  synchrotron self-absorption was probably not a factor in the rise to
maximum and that another process, likely free-free absorption, dominated.

\acknowledgments
PC thanks Jayaram Chengalur, Nissim Kanekar  and Gulabchand Dewangan for useful discussions. 
 The National Radio Astronomy Observatory is a facility of 
the National Science Foundation operated under cooperative 
agreement by Associated Universities, Inc
Support for this work was provided by the National Aeronautics and Space Administration through 
Chandra Awards  GO0-211080X,  GO2-13082X, and GO4-15065X issued by the Chandra X-ray Observatory Center, 
which is operated by the Smithsonian Astrophysical Observatory for and on behalf of the National 
Aeronautics Space Administration under contract NAS8-03060.
The research by CF is supported by the Swedish Research Council and Swedish National Space Board.  
This work made use of data supplied by the UK Swift Science Data Centre at the University of Leicester.

{\it Facilities:} \facility{Karl G. Jansky Very Large Array}, \facility{Chandra}, 
\facility{XMM-Newton}, \facility{Swift}, \facility{NuSTAR}.

\end{document}